\documentclass[12pt,preprint]{aastex}

\newcommand{\simge}{\mbox{$\stackrel{>}{_{\sim}}$}}
\newcommand\arcdegg{\hbox{$^\circ$}}

\newcommand{\micronn}{\,\hbox{$\mu$m}}
\newcommand\etal{ {\em et~al.\/}\thinspace}

\shorttitle{WR Sizes}
\shortauthors{Monnier \etal}

\begin{document}

\title{The Keck Aperture Masking Experiment: Near-Infrared Sizes of Dusty Wolf-Rayet Stars}


\author{J. D. Monnier\altaffilmark{1}, P. G. Tuthill\altaffilmark{2},
W. C. Danchi\altaffilmark{3}, N. Murphy\altaffilmark{1,4} and
T.~J.~Harries\altaffilmark{5}
}


\altaffiltext{1}{University of Michigan, Department of Astronomy, Ann Arbor, MI 48109}
\altaffiltext{2}{University of Sydney, Australia}
\altaffiltext{3}{NASA-Goddard Space Flight Center}
\altaffiltext{4}{Now at the University of Wisconsin, Department of Astronomy}
\altaffiltext{5}{University of Exeter, England}


\begin{abstract} 
  We report the results of a high angular resolution near-infrared
  survey of dusty Wolf-Rayet stars using the Keck-1 Telescope, including
  new multi-wavelength images of the pinwheel nebulae WR~98a, WR~104,
  and WR~112.  Angular sizes were measured for an additional 8 dusty
  WR stars using aperture masking interferometry, allowing us to probe
  characteristics sizes down to $\sim$20~milliarcseconds ($\sim$40~AU
  for typical sources).  With angular sizes and specific fluxes, we
  can directly measure the wavelength-dependent surface brightness and
  size relations for our sample. We discovered tight correlations of
  these properties within our sample which could not be explained by
  simple spherically-symmetric dust shells or even the more realistic
  ``pinwheel nebula'' (3-D) radiative transfer model, when using
  optical constants of Zubko.  While the tightly-correlated surface
  brightness relations we uncovered offer compelling {\em indirect}
  evidence of a shared and distinctive dust shell geometry amongst our
  sample, long-baseline interferometers should target the
  marginally-resolved objects in our sample in order to conclusively
  establish the presence or absence of the putative underyling
  colliding wind binaries thought to produce the dust shells around WC
  Wolf-Rayets.

\end{abstract}




\section{Introduction}
The existence of dust shells around Wolf-Rayet stars has long posed a
mystery: how can dust nucleate and survive near these hot stars in the
presence of harsh ultraviolet radiation and gas densities lower than
those found in other dust forming environments (e.g., AGB winds)?  While
novel chemical pathways for producing special carbonaceous grains have
been proposed \citep{zubko98,cherchneff2002}, recent theoretical and
observational results point strongly to a very different solution.

Inspired by the periodic dust formation episodes around the colliding
wind binary WR~140 \citep{moffat87, williams90}, \citet{usov91}
suggested that dust formation might be catalyzed in the compressed
layer of gas at the interface of colliding winds in WR+OB binary
systems.  This hypothesis was confirmed by the surprising imaging of a
``pinwheel'' nebula dust spiral around WR~104 \citep{tuthill99}.
Indeed, \citet{monnier99} went so far as to suggest that perhaps {\em
  all} dusty WR systems are hiding colliding-wind binary systems
\citep[see also][]{dw2000} and this was further supported by the
discovery of non-thermal radio emission around WR~104, WR~98a, and
WR~112 \citep{monnier2002c}.

The previously-cited imaging work was carried out using the
diffraction-limited capabilities of the world's largest optical telescope at the
Keck Observatory.  
High-resolution imaging was achieved using aperture masking
interferometry, whereby the Keck-1 primary mirror is converted to a
VLA-style interferometric array \citep{tuthill2000} of many subapertures.
This technique has been shown to be superior than current adaptive optics
systems for high-fidelity imaging and size estimations for 
marginally-resolved objects \citep{rajagopal2004}.

Here we report an extension to the initial imaging work of WR 104 and
WR 98a. We observed all WR stars (accessible from Mauna Kea) brighter
than $m_K$ 6.5 in order to investigate the binary hypothesis as to the
origin of dust production.  Although imaging was only possible for one
additional source (WR~112), we report the characteristic size
measurements for a total of 11~sources and discuss the significance of
these findings.

\section{Observations}
Our group has been carrying out aperture masking interferometry at the
Keck-1 telescope since 1996.  We have published images and size
measurements with unprecedented angular resolution on topics ranging
from young stellar objects, carbon stars, red supergiants, and
photospheric diameters of Mira variables
\citep[e.g.,][]{monnier99a,tuthill2000b,tuthill2000a,danchi2001}.  A
full description of this experiment  can be found in
\citet{tuthill2000}, with further discussion of systematic errors in
\citet{monnier2004a}.

The NIRC camera with image magnifier \citep{matthews96} was used in
conjunction with the aperture masking hardware. For this work, we used
an aperture mask with an annulus that is 8-m in diameter as projected
onto the Keck primary. This mask gives us complete UV coverage and
sensitivity for targets brighter than $m_K \sim$ 6 with a modest loss
in calibration precision due to additional redundancy noise \citep[for
further discussion, see][]{tuthill2000}.  The data frames were taken
in speckle mode ($T_{\rm int}$=0.14\,s) to freeze the atmosphere.  A
variety of filters were used and the characteristic center wavelengths
and widths can be found in Table\,\ref{filters} (spectral scans can be
found in Keck-NIRC users manual).

Here we report the full body of Wolf-Rayet size measurements collected
during the entire period of the Keck aperture masking experiment. A
target list and observing log can be found in Tables\,\ref{targets} \&
\ref{observations}.  All V$^2$ and closure phase data are available
from the authors;
all data products are stored in the FITS-based, optical interferometry
data exchange format (OI-FITS), recently described in
\citet{pauls2005}.

\section{Determining Characteristic Sizes}

\subsection{Methodology}
While the basic data reduction has been described in previous
papers \citep{tuthill2000}, this paper is the first Keck aperture
masking paper to explicitly deal with a sample of partially-resolved
objects.  This is important here since we intend
to measure the 2-dimensional size and shape of Wolf-Rayet dust shells
in order to search for asymmetries.  If we measure an ellipticity in
our data, what confidence do we have that an elongation is not due to
poor calibration of the optical transfer function?  To answer this
question we have undertaken a systematic investigation of
our calibration errors.

The errors for this experiment are
dominated by statistical and seeing calibration errors.  The
statistical error arises from the contribution of photon noise and
read noise to our measurement process.  Calibration error arises
because the optical transfer function varies with time and telescope
pointing.  In order to correct for this latter effect, we always
observe a point-source reference star nearby in time and angle from
our target.  In the process of calibration, 
raw visibilities from the target are divided by those from the calibrator.
Hence, calibration error is
multiplicative and affects high visibility data the most in absolute
terms ($\Delta V$).  For well-resolved objects, these two types of error
are comparable.  For small objects ($\lesssim 25$~mas) calibration
error is dominant and limits our ability to say whether an object is
resolved.

\subsubsection{Functional Form and Statistical Errors}
As part of our analysis, the calibrated visibility data 
are fit to the following generic function (based on a 2-dimensional Gaussian):
\begin{eqnarray}
V^2 = [V_o\,e^{-(\alpha u^2 + \beta v^2 + \gamma uv)} + V_P]^2,
\label{funct-form1}
\end{eqnarray}
where $V_P$ is constant and represents a point source contribution
(which was set to zero for this study).  The parameters $\alpha$,
$\beta$, and $\gamma$ are not constrained to be positive, thus allowing the
function to curve upwards in cases where calibration errors are extreme.

Under a rotation through angle $\phi$, this function can be projected onto the major and minor axes ($u',v'$), 
written in the form

\begin{eqnarray}  
V^2 =
\left[V_o\,e^{-4\ln{2}\left[\left(\frac{u'}{u'_{\rm{{FWHM}}}}\right)^2 +
\left(\frac{v'}{v'_{\rm{FWHM}}} \right)^2\right]} + V_P\right]^2
\label{funct-form2}
\end{eqnarray}
with the parameters of this fit being given by

\begin{eqnarray}
\phi &=& \frac{1}{2} \arctan{\left(\frac{\gamma}{\alpha-\beta}\right)}\\
(u'_{\rm{FWHM}})^2&=&\frac{4\ln{2}\cos{2\phi}}{\alpha-(\alpha+\beta)
  \sin^2{\phi}} \\
(v'_{\rm{FWHM}})^2&=&\left[\frac{\alpha+\beta}{4\ln{2}}-
\frac{1}{(u'_{\rm{FWHM}})^2} \right]^{-1}
\label{funct-temp}
\end{eqnarray}

These expressions can be converted to the angular FWHM  along the major/minor axis
using the formula FWHM$_{\rm Major}=\frac{4\ln{2}}{\pi u'_{\rm{FWHM}}}$ and FWHM$_{\rm{Minor}}=\frac{4\ln{2}}{\pi v'_{\rm{FWHM}}}$

Statistical errors are dealt with by using bootstrap sampling
\citep{bootstrap} with the function given in
Equation~\ref{funct-form1}.  The fitting
routine \begin{tt}MPFIT\end{tt} in Craig
Markwardt's \begin{tt}MINPACK\end{tt}\footnote{See http://cow.physics.wisc.edu/$\sim$craigm/idl/} suite of IDL programs is used as
the basis for the bootstrap.

\subsubsection{Monte-Carlo Method for Determining Calibration Error}
\label{mcerrors}
Quantifying the magnitude of calibration errors in this experiment has been
difficult.  Due to the very limited time available on a Keck telescope, the
number of stars used as calibrators for a given filter is usually small,
ranging from less than 5 up to perhaps a dozen for the most commonly used bands.
While precise systematic errors are difficult to determine due to small
number statistics, it is possible to estimate the magnitude of the calibration
accuracy.

The first step in the analysis of calibration errors is to
cross-calibrate all of the calibration stars of a given night
(with same bandpass filter and aperture mask) against each other and
fit the resulting visibility curves with the quadratic function
introduced in the last section (Eq.\ref{funct-form2}).  At this point, the
calibrator-calibrator pairs are weighted according to proximity to
each other in time and elevation according to $e^{-\left( \frac{\Delta
      t}{\Delta t_*}\right)^2 - \left( \frac{\Delta
      \theta_{el}}{\Delta \theta_{el*}}\right)^2}$ with $\Delta t_*
\sim 2000$~s and $\Delta\theta_{el*} \sim 20^\circ$.

In order to propagate calibration errors into the errors on the
Gaussian FWHM sizes that we wish to measure, we start with a
visibility curve with a known FWHM (using the same baselines as the
Keck data).  Then we multiply this perfect simulated data by various
calibrator-calibrator pair data and then re-calculate a size using the
quadratic model developed above.  This simulates what the measured visibility would
look like for a star of a known size using different calibrators.  By
Monte Carlo sampling over many cal-cal pairs according the weights
described above, we build up an estimate of the likely error in the
final size measurement.

\begin{figure}

\begin{center}
\includegraphics[angle=90,scale=0.25]{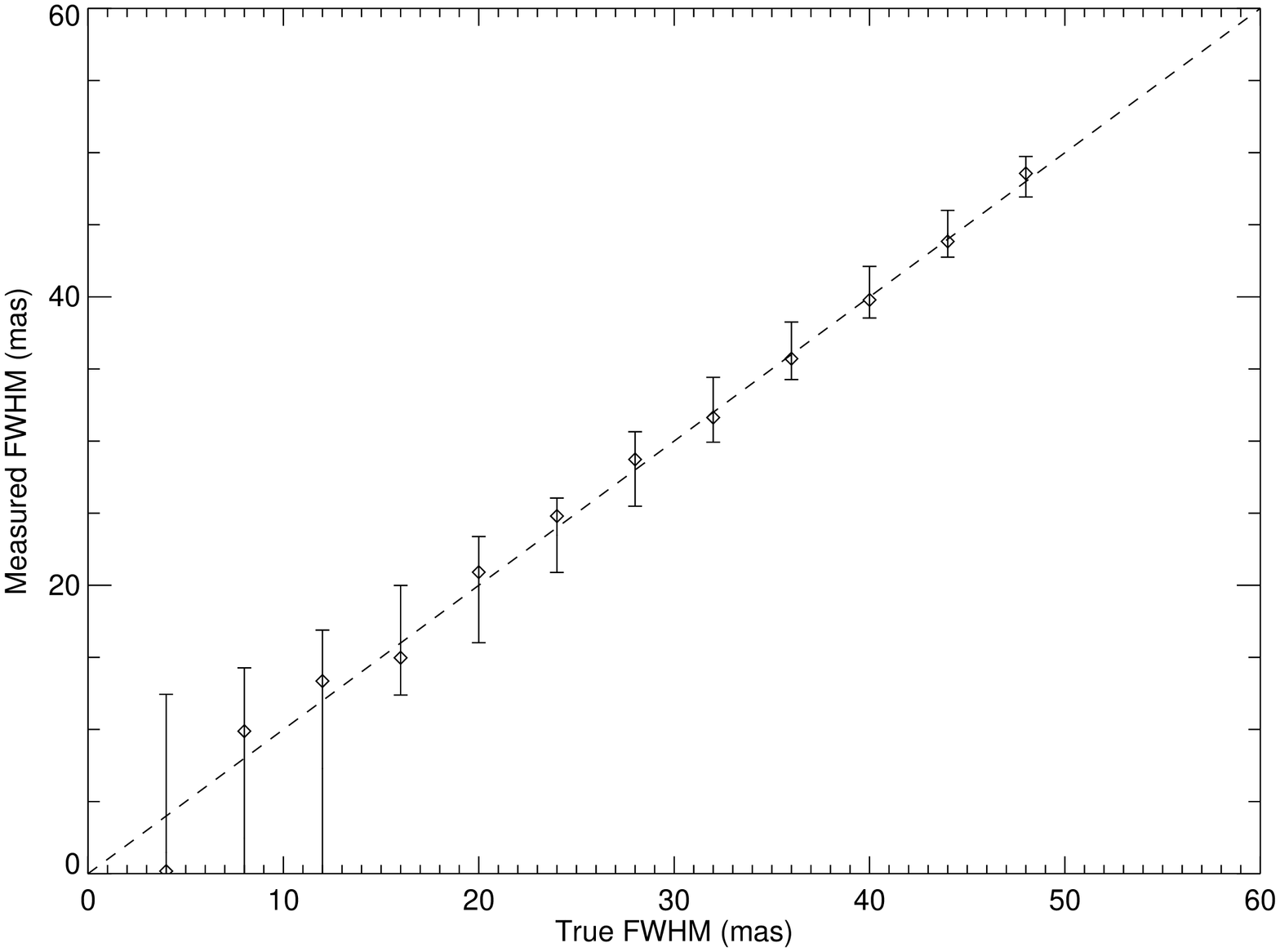}~~~\\
\includegraphics[angle=90,scale=0.25]{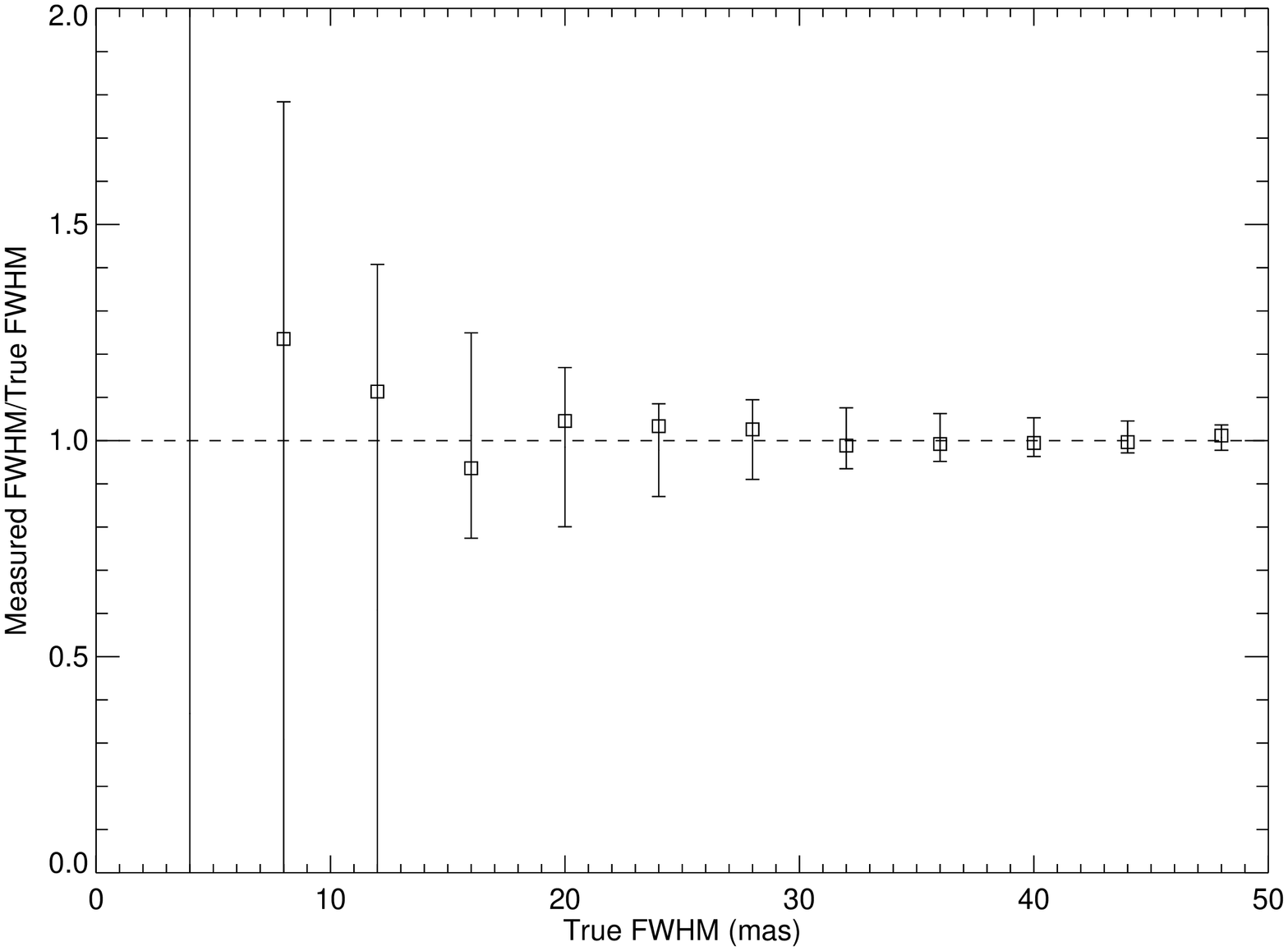}~~~\\
\includegraphics[angle=90,scale=0.25]{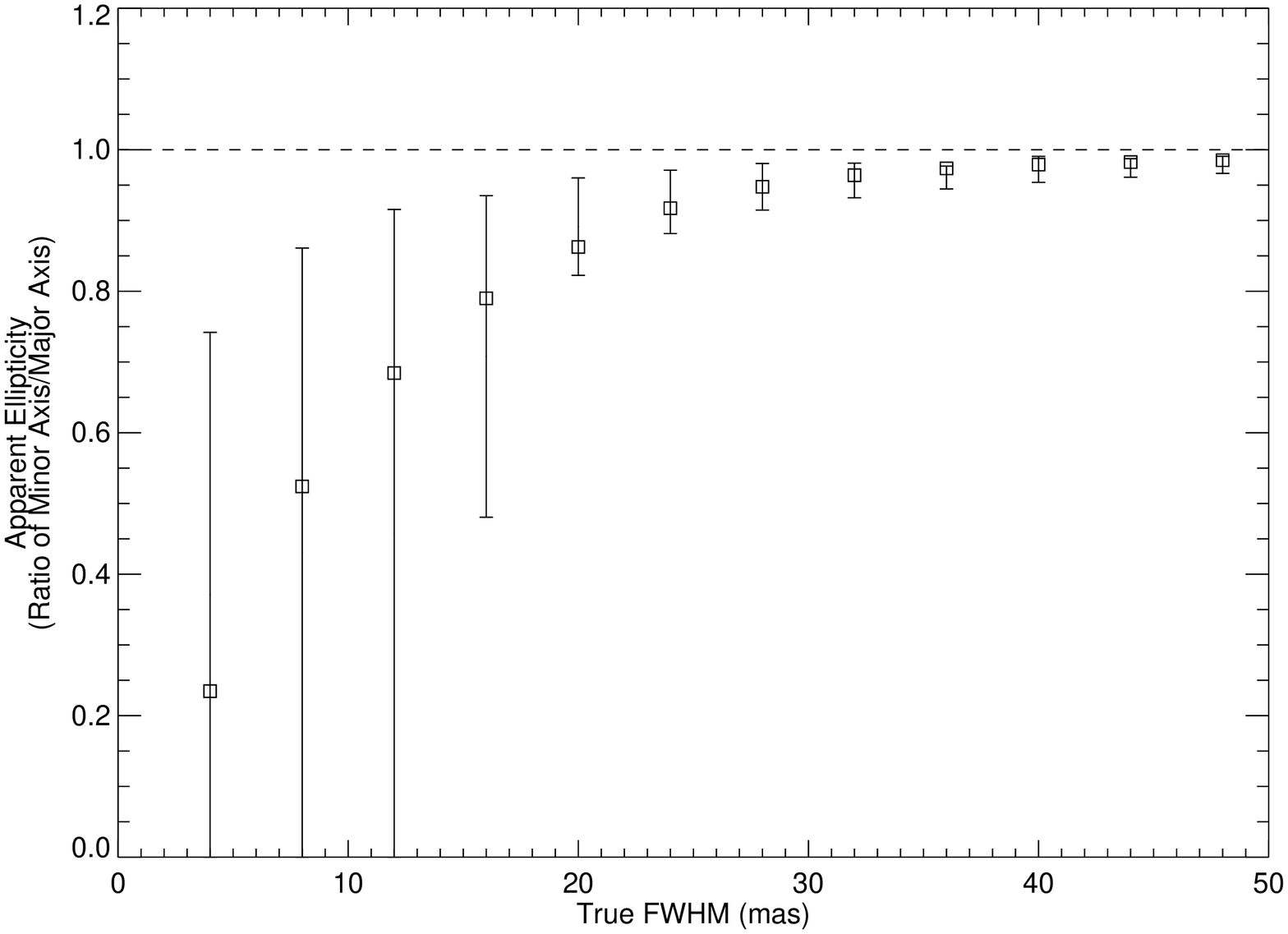}~~~
\caption{These graphs show the results for the fitting of 
simulated data using calibrators from the June 2000 epoch for the 
CH4 band using the Annulus mask. The top panel shows the measured Gaussian FHWM from
the simulated data compared to the input model. The middle panel re-displays this
information as a ratio. The bottom panel explores how miscalibration induces
ellipticity into the fits, even for an input circular Gaussian.
The dashed lines indicate the position of the
original simulated data.    All error bars indicate
1-$\sigma$ certainty and data points indicate medians.\label{testdata}
}
\end{center}
\end{figure}

The results are summarized in Figure~\ref{testdata} for the 2000 June
epoch in the CH4 filter using the Annulus mask.  The most important
feature of these plots is determination of the effective angular
resolution limit of this experiment.  We see that simulations of an
unresolved point source, when passed through the calibration study
described above, can yield apparent sizes as large as
12~milliarcseconds (Gaussian FWHM).  From this, we conclude that our
resolution limit corresponds to Gaussian FWHMs of 12-15 mas
($\lambda\sim2.2\mu$m), approximately 4$\times$ better than the formal
Keck diffraction limit $1.22 \lambda / D \sim 55~$mas.  
For nights of poor seeing, our size upper limit on FWHM increases to
15-20~mas.

As stated before, we would like to use the Keck masking data to
measure modest asymmetries in the dust shells of Wolf-Rayet stars,
since this can tell us about the dust production mechanism
(single-star vs binary interaction). In order to test our sensitivity
to asymmetries, we fitted 2-D Gaussians to our calibrator study data
and these results are also shown in Figure~\ref{testdata}.  Our study
shows that if we impose typical calibration errors on a circularly symmetric
object with FWHM 20~mas, we will regularly find asymmetries with
measured ratio FWHM$_{\rm minor}$ / FWHM$_{\rm major} =$ 0.8.  Thus,
except for extremely asymmetric objects, any study on asymmetries from
this data will have to restrict itself to objects that are larger than
about $\simge$25-30~mas.

We note that the calibration errors determined by this method are
over-estimates, since typical calibrator-calibrator pairs suffer from
greater time difference and angular distance than target-calibrator
pairs used for actual science measurements (this is confirmed for a
few objects where we have multiple independent data sets).  Thus we
can assume that the errors calculated in this section are conservative
and  the true calibration errors
should be up to a factor of 1.5 to 2 smaller for nights of comparable
seeing.  

The greatest uncertainty in this analysis is that it uses a very
limited number of calibrator stars spread out over a night.  Our
understanding of the errors would improve greatly if a series of
identical exposures of a calibrator star were to be taken
sequentially\footnote{The Keck Time Allocation Committee is unlikely to give
time to such a proposal}.  This would allow us to see the short timescale
variations in a calibrator's visibility curve, and thus how much
variance there is in the calibration process.  If very large
variations in the quadratic fits remain on timescales as short as
this, the calibration process may need further refinement.

\begin{figure}

\begin{center}
\includegraphics[angle=90,scale=0.25]{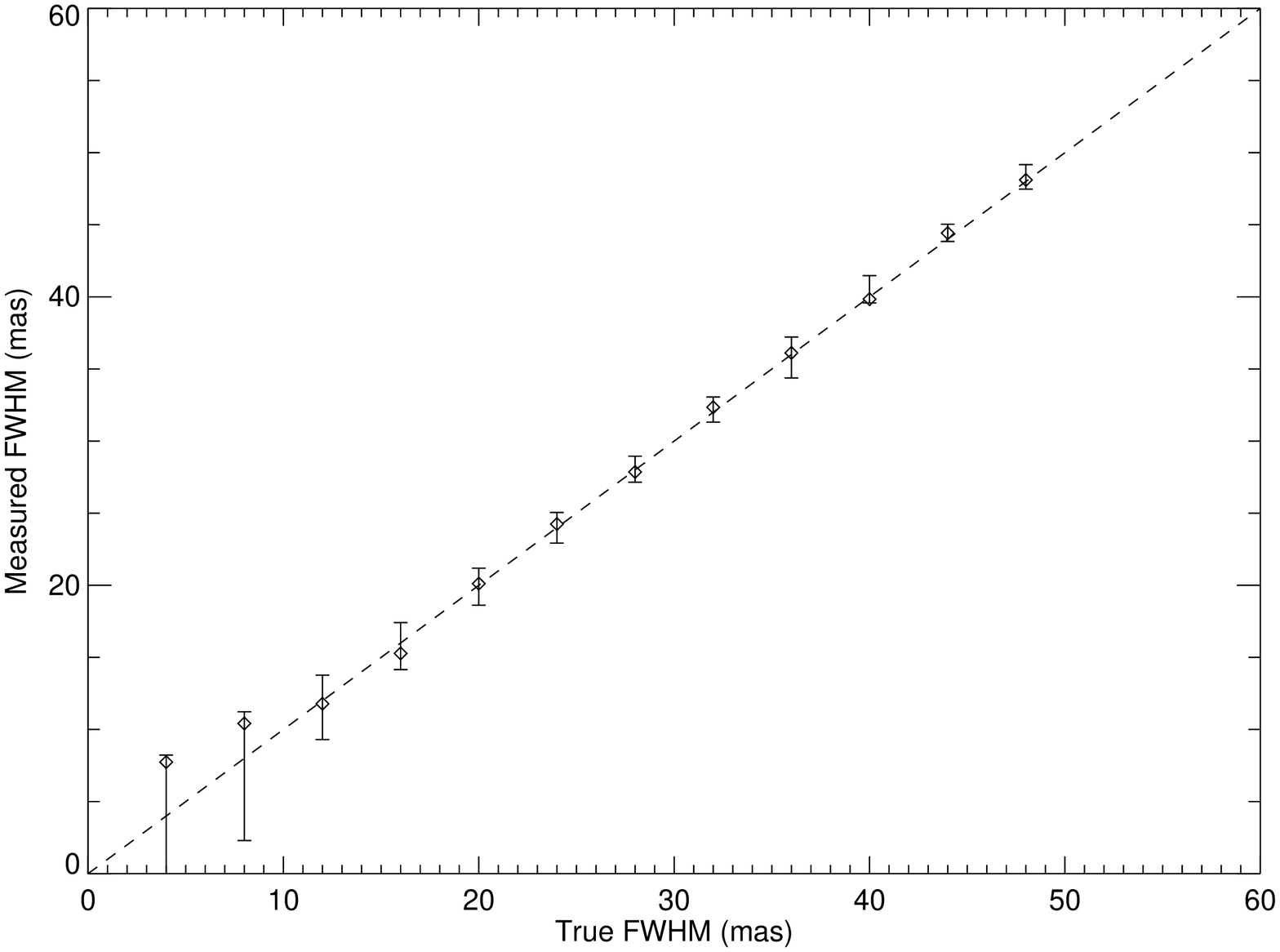}~~~\\
\includegraphics[angle=90,scale=0.25]{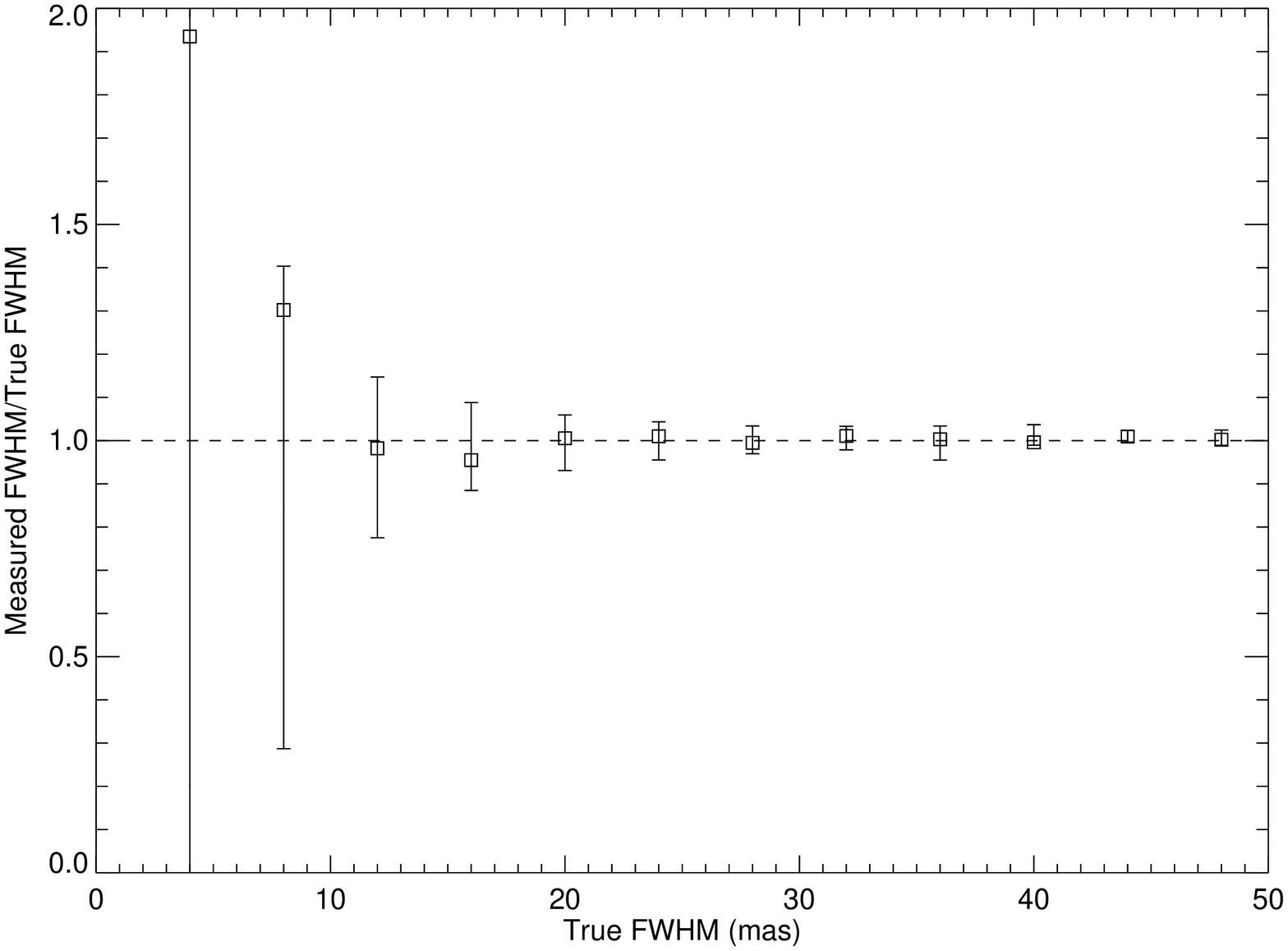}~~~\\
\includegraphics[angle=90,scale=0.25]{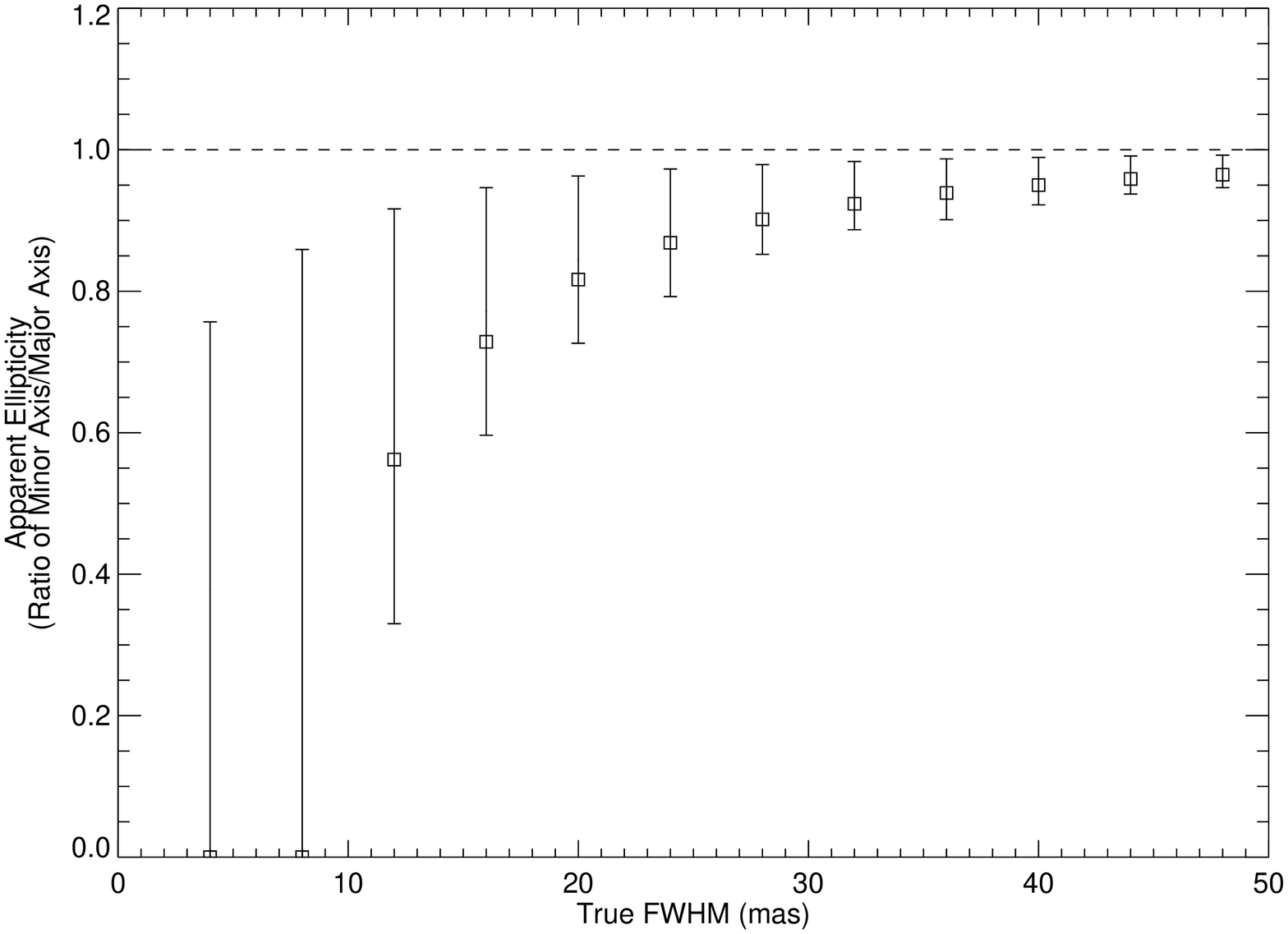}~~~
\caption{These graphs show the results for the fitting of recalibrated
simulated data using calibrators from the June 2000 epoch for the 
H band using the Annulus mask. The format of the plots are the same as previous figure.
\label{testdataH}
} 
\end{center}
\end{figure}

\begin{figure}

\begin{center}
\includegraphics[angle=90,scale=0.25]{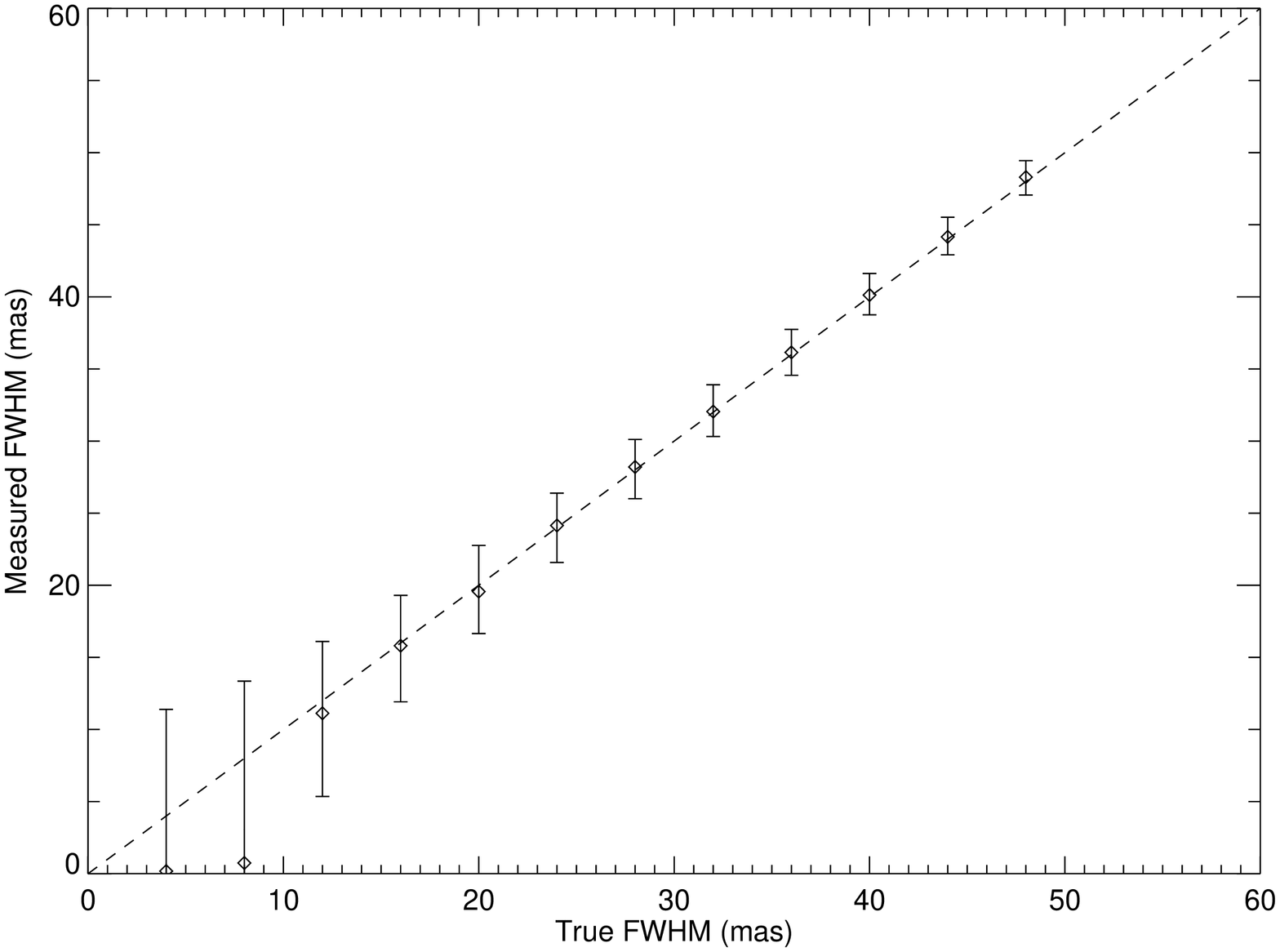}~~~\\
\includegraphics[angle=90,scale=0.25]{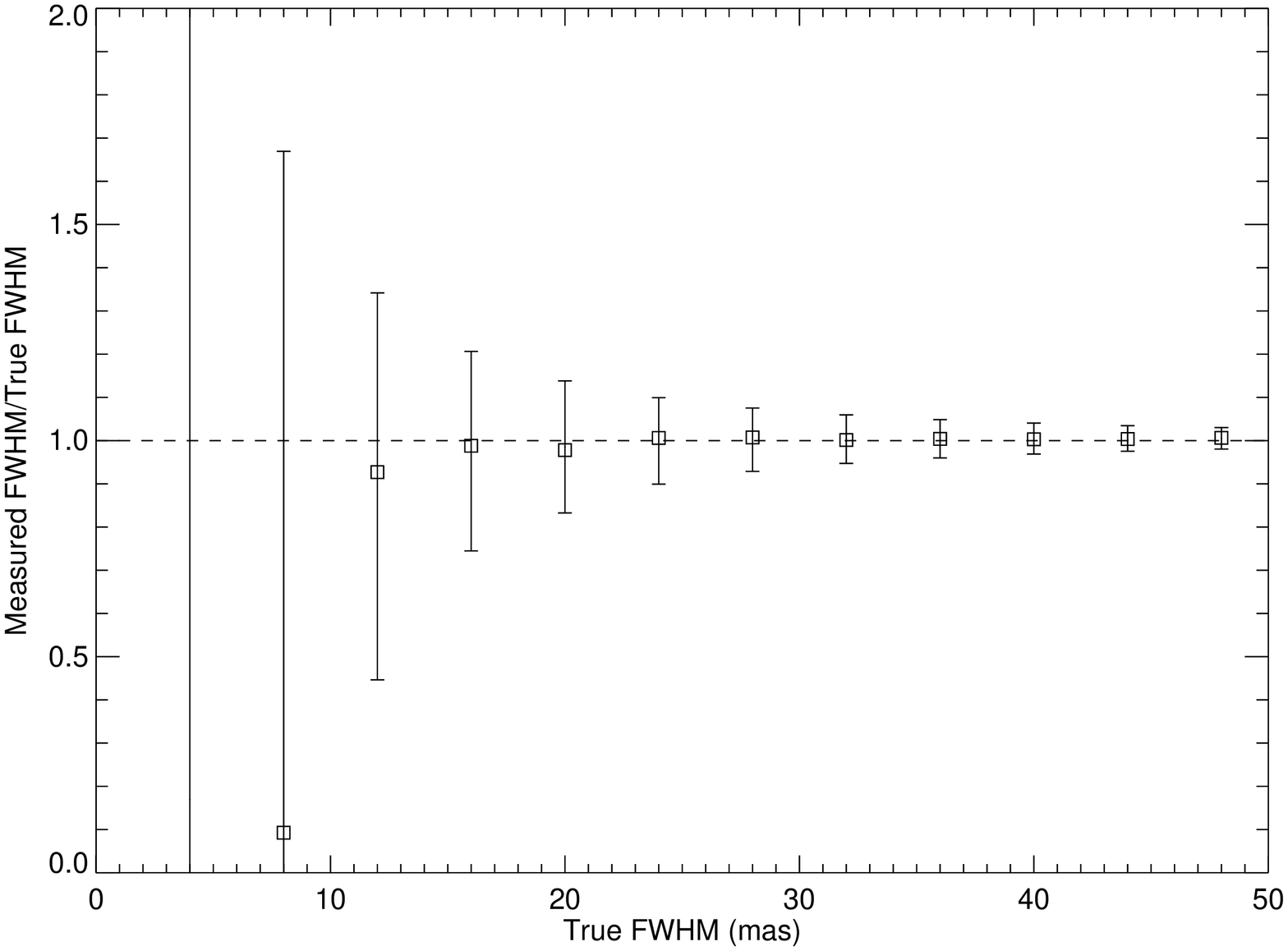}~~~\\
\includegraphics[angle=90,scale=0.25]{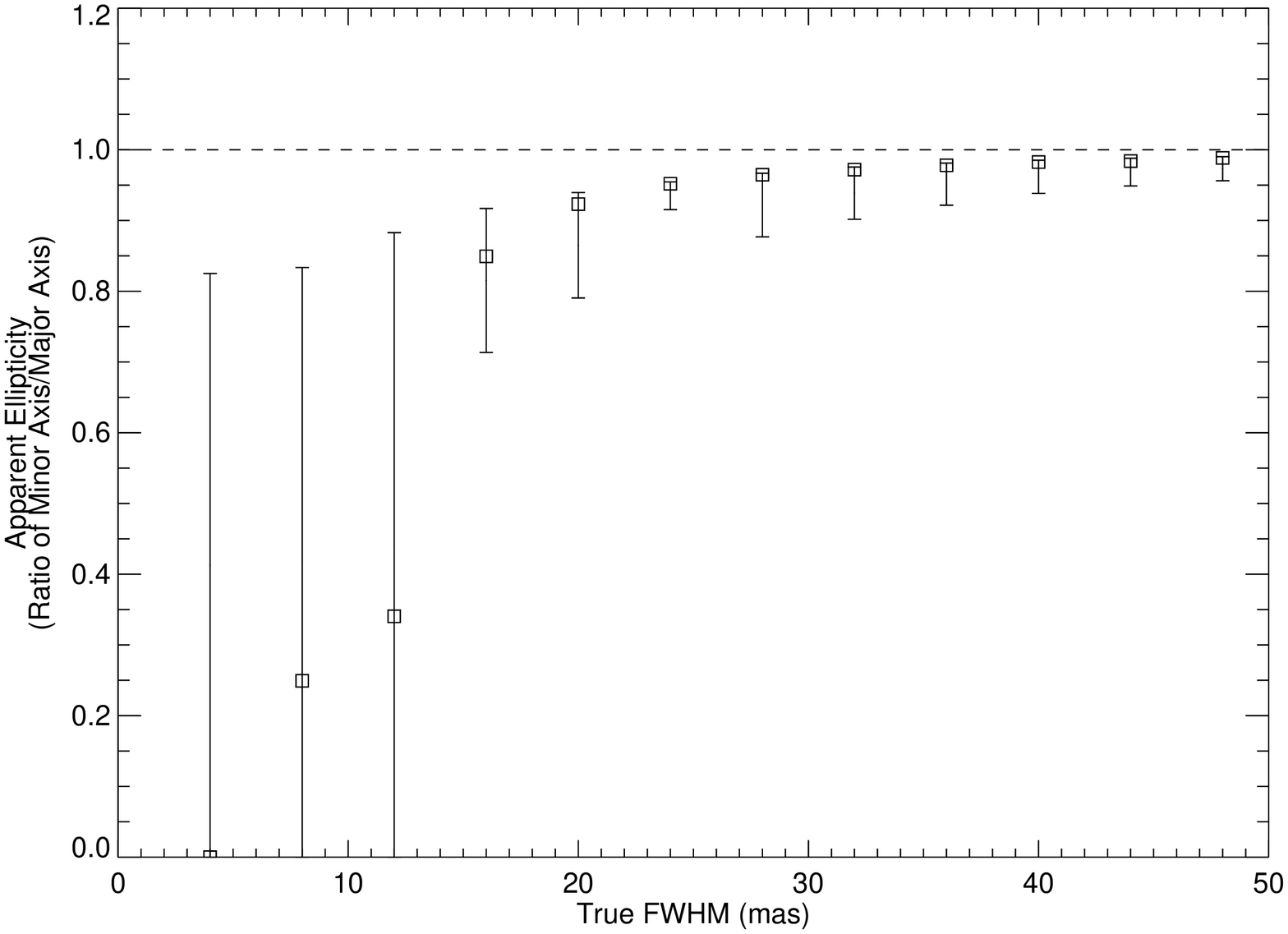}~~~
\caption{These graphs show the results for the fitting of recalibrated
simulated data using calibrators from the June 2000 epoch for the 
PAHcs  band using the Annulus mask. The
format of the plots are the same as previous figure.
\label{testdataPAHcs}
} 
\end{center}
\end{figure}

The same study has been done for H band ($\lambda=1.65\mu$m) and PAHcs
($\lambda=3.08\mu$m), and these results are also included (in the same
format) in Figures~\ref{testdataH} \& \ref{testdataPAHcs}.

\section{Results}

\subsection{Multi-wavelength Characteristic Sizes}
Circularly-symmetric gaussian emission profiles were fitted to the
azimuthally-averaged visibility data for all targets and the results 
can be found in Table~\ref{results1}.  The uncertainty estimates were
based on the combined error from statistical variations (using
bootstrap) and from the calibration errors discussed in the last
section.  Calibration error is typically the dominant error for these
data.

The largest dust shells (WR~48a, WR~98a, WR~104, WR~112, and WR~140)
show strong deviations from the Gaussian at the longer Keck baselines.
This is not at all surprising since WR~98a, WR~104, WR~112 and WR~140
are all known to have spiral and/or asymmetric dust shells
\citep{tuthill99,monnier99,rr99,monnier2002b,marchenko2002}.  It is likely that the
other smaller dust shells in our sample are also not simple Gaussians,
although we were unable to detect deviations due to insufficient
angular resolution (all marginally resolved objects look Gaussian-like
to an interferometer).  

\begin{figure}

\begin{center}
\includegraphics[width=5in]{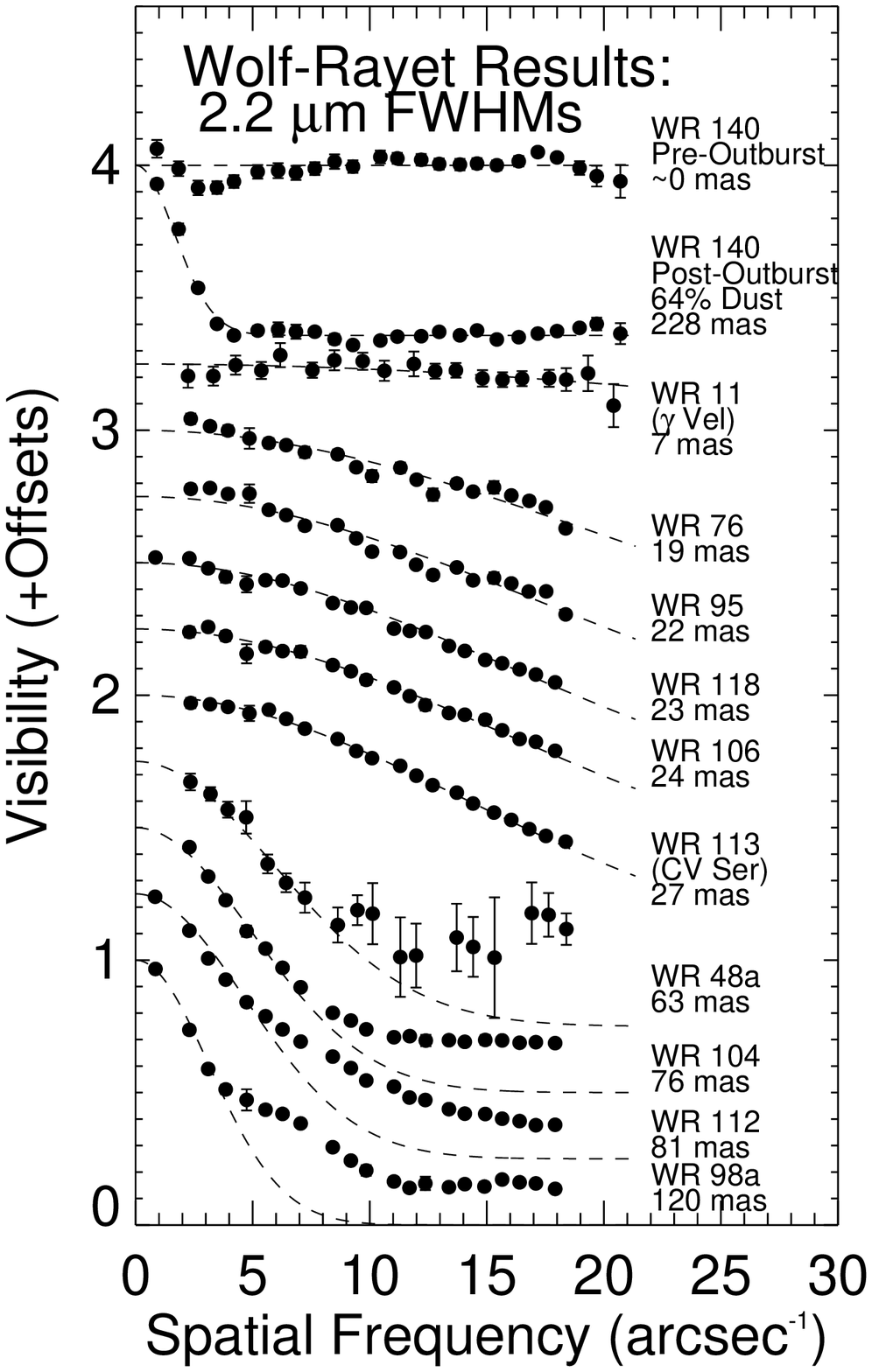}
\caption{This figure shows the azimuthally-averaged visibility curves at 2.2$\mu$m 
for Wolf-Rayet
stars in our sample, along with best-fit Gaussian curves. Each visibility curve is
offset by 0.25 from the next. For the most resolved dust shells, the Gaussian curves
were only fitted to the ``large-scale'' component of the visibility data ($V>0.5$).
See Table~\ref{results1} for detailed fitting results.\label{Ksizes}
} 
\end{center}
\end{figure}

\begin{figure}

\begin{center}
\includegraphics[width=5in]{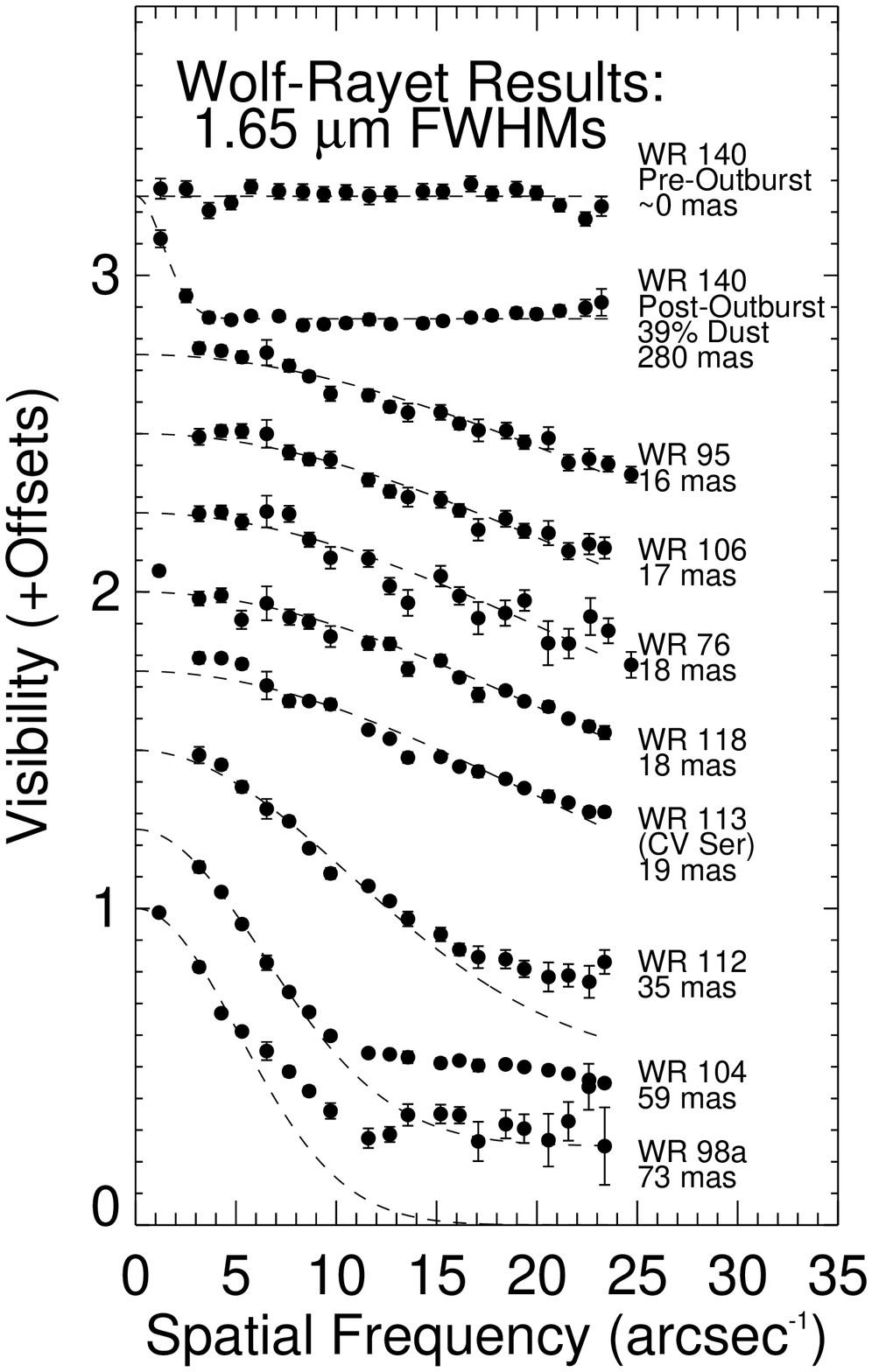}
\caption{This figure shows the azimuthally-averaged visibility curves at 1.65$\mu$m 
for Wolf-Rayet
stars in our sample, along with best-fit Gaussian curves. Each visibility curve is
offset by 0.25 from the next. For the most resolved dust shells, the Gaussian curves
were only fitted to the ``large-scale'' component of the visibility data ($V>0.5$).
See Table~\ref{results1} for detailed fitting results.\label{Hsizes}
} 
\end{center}
\end{figure}

\begin{figure}

\begin{center}
\includegraphics[width=5in]{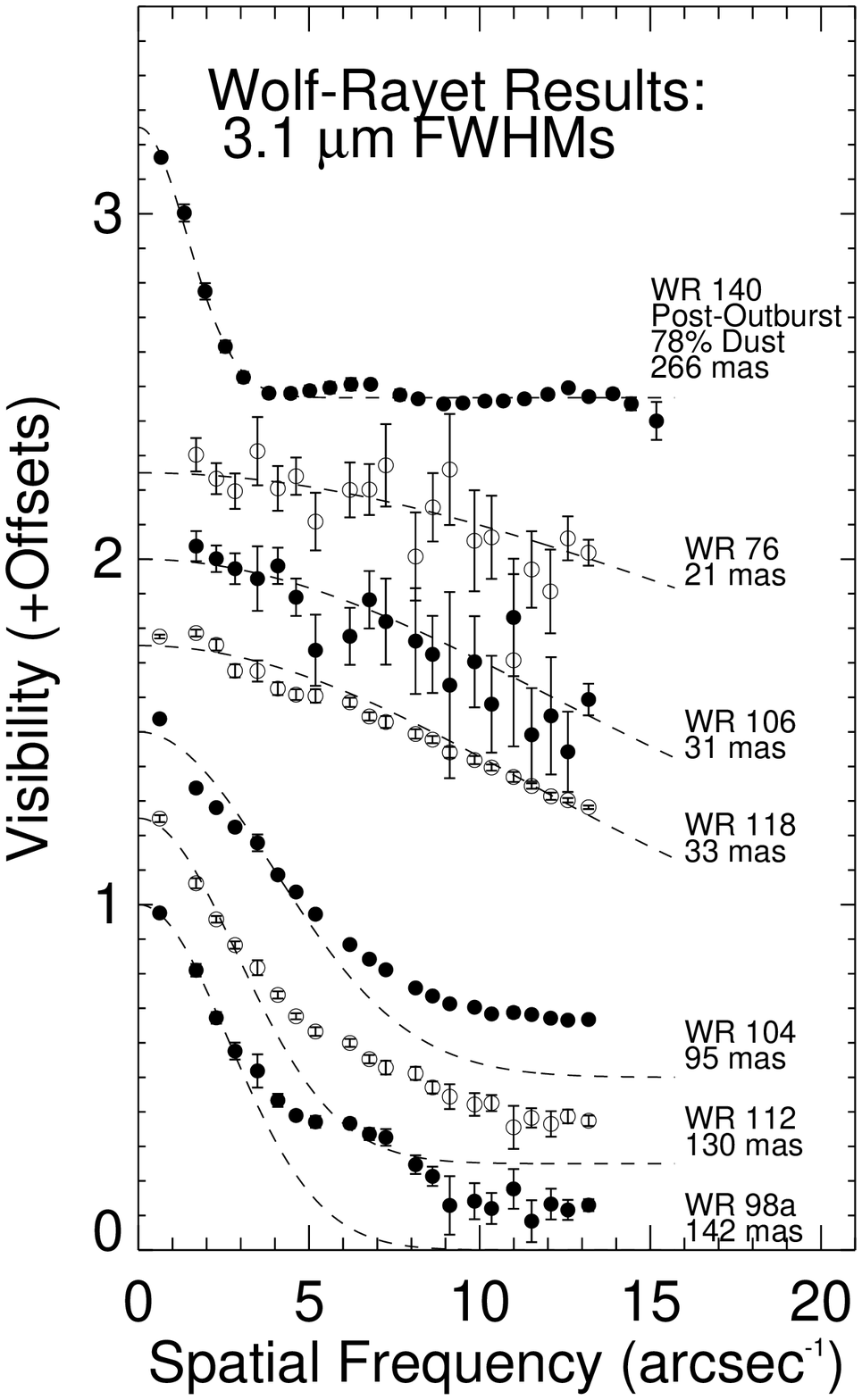}
\caption{This figure shows the azimuthally-averaged visibility curves at 3.1$\mu$m 
for Wolf-Rayet
stars in our sample, along with best-fit Gaussian curves. Each visibility curve is
offset by 0.25 from the next. For the most resolved dust shells, the Gaussian curves
were only fitted to the ``large-scale'' component of the visibility data ($V>0.5$).
See Table~\ref{results1} for detailed fitting results.\label{Psizes}
} 
\end{center}
\end{figure}

In order to calculate a consistent
``characteristic size'' for all targets, we only fit to short-baseline
data where the visibility is greater than 0.5.  By fitting only to
short-baseline, high-visibility data, the reported characteristic
sizes represent the overall scale of emission and are not strongly
affected by the spiral and/or small-scale structures of the nebulae,
if present.  Our results are graphically shown for 2.2$\mu$m,
1.65$\mu$m and 3.08$\mu$m in Figures~\ref{Ksizes}-\ref{Psizes}.

Subject to the (large) uncertainties in ellipticity discussed in \S\ref{mcerrors},
we did not find evidence for elongation in the emission except for the cases of 
the largest dust shells, WR~98a, WR~104, and WR~112.  For these, we have
reported the mean Gaussian FWHM and the ratio of the minor axis to the major axis.
Just as for the characteristic sizes discussed above, this ellipticity parameter only
refers to the ``large-scale'' emission component of the dust shell and 
not the smaller-scale details, such as the inner spiral windings of WR~104.

\subsection{Correction for Stellar Contribution}
\label{corrections}
While for most stars the near-infrared emission is completely
dominated by dust emission, some targets in our sample have
significant stellar contributions.  By carrying out simple spectral
energy density (SED) fitting, we can estimate the fraction of light
coming from the star (compared to dust) and apply a correction
factor. Essentially the visible photometry is fit with a Kurucz model
allowing the stellar size and reddening to vary; the stellar IR flux 
can then be estimated by extrapolation.
With this knowledge, we can  more accurately estimate the true size of
the dust shell (not the star+dust emission).  
This is a common
procedure in other areas of astronomy, such as estimating the sizes of
disks around young stellar objects, and we follow 
procedures documented  elsewhere
\citep[e.g.,][]{rmg2001,monnier2005}.

This correction is strongest for $\gamma$~Vel (WR~11) which does not
show much infrared excess, but since the observed characteristic size
was consistent with a point source, we do not apply the correction.
For cases where the correction is more than 5\% (WR~48a, WR~95,
WR~98a, WR~106, WR~113), we have included the new size estimates (and the 
estimates for the dust fraction) in
Table~\ref{results2}.  We estimate the correction factor is only known
to 50\% due to difficulty in uniquely fitting the broadband SEDs,
and this error has been included in the corrected sizes in
Table~\ref{results2}.

Note that the post-outburst WR~140 dust shell is so large that it was
obviously necessary to correct for the stellar contribution (see
top curves of Figure~\ref{Ksizes}) and this was
done for Table~\ref{results1}.  The WR~140 dust shell was considered in detail in
\citet{monnier2002b} and we refer the reader to this paper for further discussion on this
object.

\subsection{Aperture Synthesis Imaging}

Three systems in our sample are sufficiently resolved that they can be
imaged using aperture synthesis techniques.  The Maximum Entropy
Method (MEM) \citep{sb84,mem83} has been used to create
diffraction-limited images from the interferometric data, as
implemented in the VLBMEM package by \citet{sivia87}.  

\begin{figure}

\begin{center}
\includegraphics[width=6in]{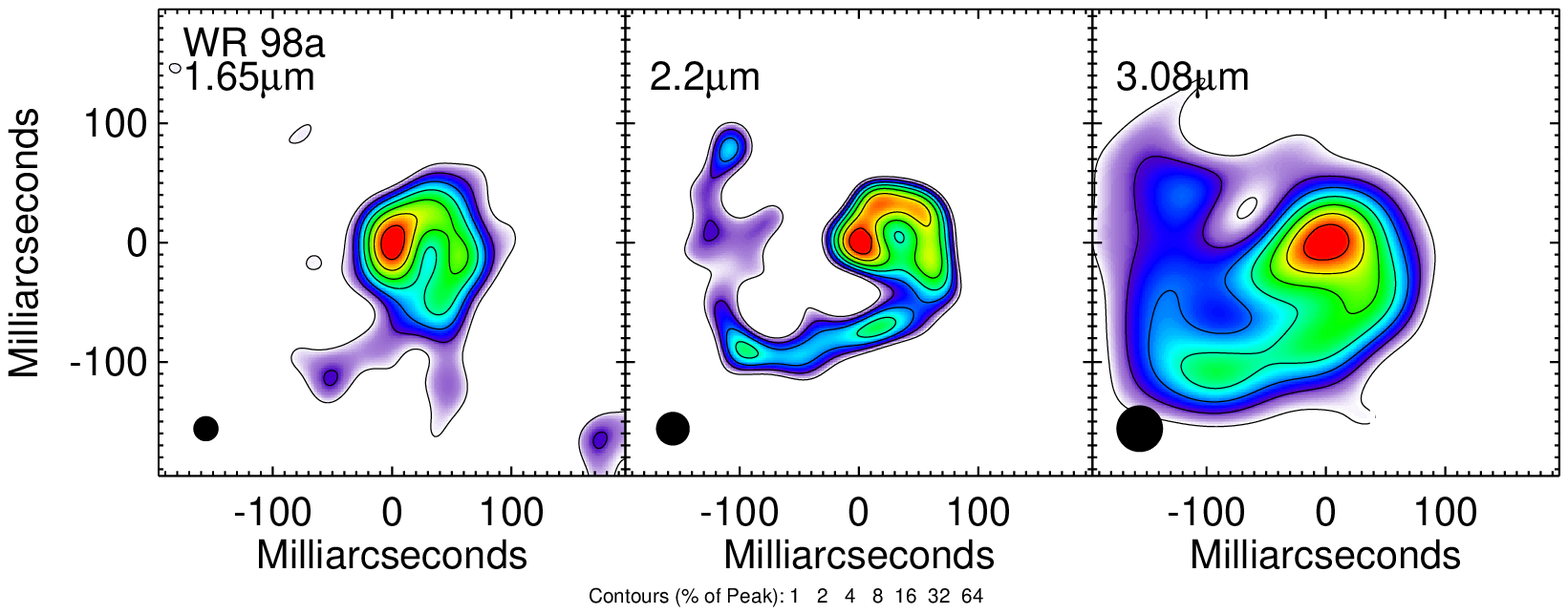}
\caption{Here we show multiwavelength aperture synthesis images of WR~98a on UT 2000 Jun 24, at
1.65$\mu$m, 2.2$\mu$m, and 3.08$\mu$m.  Note that the resolution is lower at the
longer wavelengths, as indicated by the ``beam'' spot located in the bottom-left
corner of each panel (representing the best achievable angular resolution, $\Delta\Theta =
\frac{\lambda}{2 B_{\rm max}}$).\label{fig98a}
} 
\end{center}
\end{figure}

\begin{figure}

\begin{center}
\includegraphics[width=6in]{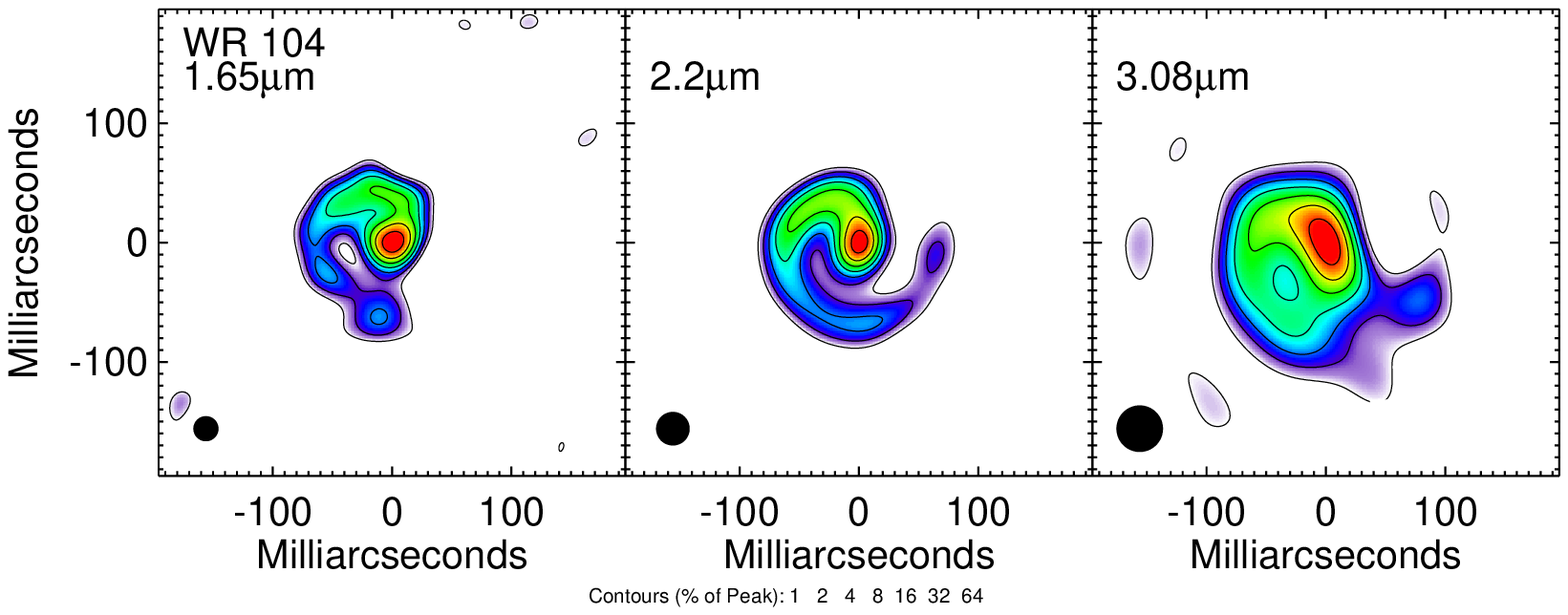}
\caption{Here we show multiwavelength aperture synthesis images of WR~104 on UT 2000 Jun 24, at
1.65$\mu$m, 2.2$\mu$m, and 3.08$\mu$m.  Note that the resolution is lower at the
longer wavelengths, as indicated by the ``beam'' spot located in the bottom-left
corner of each panel (representing the best achievable angular resolution, $\Delta\Theta =
\frac{\lambda}{2 B_{\rm max}}$).\label{fig104}
} 
\end{center}
\end{figure}

\begin{figure}

\begin{center}
\includegraphics[width=6in]{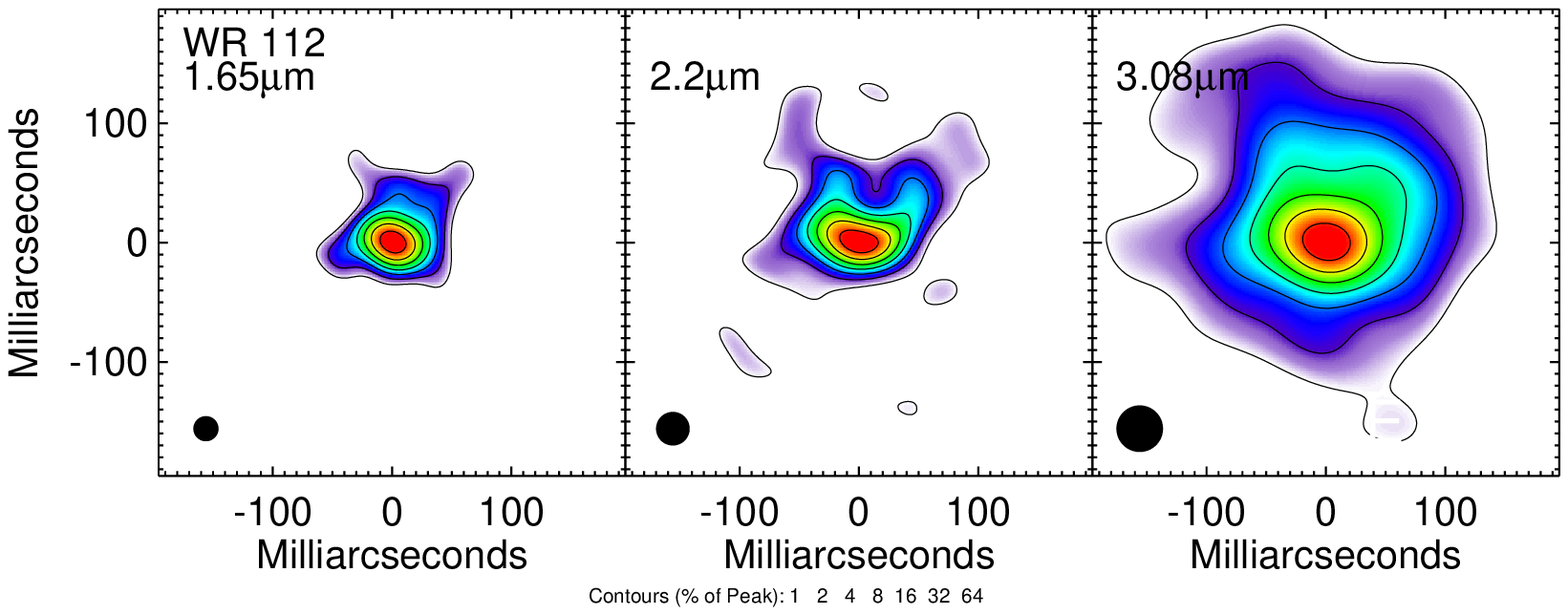}
\caption{Here we show multiwavelength aperture synthesis images of WR~112 on UT 2000 Jun 24, at
1.65$\mu$m, 2.2$\mu$m, and 3.08$\mu$m.  Note that the resolution is lower at the
longer wavelengths, as indicated by the ``beam'' spot located in the bottom-left
corner of each panel (representing the best achievable angular resolution, $\Delta\Theta =
\frac{\lambda}{2 B_{\rm max}}$).\label{fig112}
} 
\end{center}
\end{figure}

Figures~\ref{fig98a}-\ref{fig112} show new multi-wavelength images of
WR~98a, WR~104, and WR~112 at wavelengths of 1.65$\mu$m, 2.2$\mu$m,
and 3.08$\mu$m.  The resolution degrades with increasing wavelength
and we adopt an effective angular resolution of 21~mas, 28~mas, and
39~mas for 1.65$\mu$m, 2.2$\mu$m, and 3.08$\mu$m, corresponding to
$\Delta\Theta=\frac{\lambda}{2B_{\rm mas}}$.  While we have presented
images of WR~98a and WR~104 in the past, images given here are at a new
epoch and we have included additional wavelengths.

Figure~\ref{fig112} contains the first published resolved images of the
near-infrared dust shell around WR~112 \citep[preliminary 
results were shown at the ``Interacting Winds from Massive Stars'' workshop in 2000;][]{monnier2002winds}.  The asymmetric nature of this
dust shell was already discovered using lunar occultations
\citep{rr99}.  Recently, \citet{marchenko2002} reported a spiral-like
dust plume around WR~112 at {\em mid-infrared} wavelengths.
Interestingly, we do not see an obvious spiral structure in these new
near-infrared images, rather we see only evidence for filaments/arcs and
one-sided nebulosity.  

\section{Discussion}

This paper significantly expands the number of
WR stars with known angular sizes. and we wish to use this information to
probe the nature of the dusty outflows so common in WC systems.  The Wolf-Rayet systems with
the largest dust shells, WR~48a, WR~98a, WR~104, WR~112, WR~140, are
{\em all in confirmed colliding wind systems} (from detection of
non-thermal radio emission).  Unfortunately, we do
not have enough angular resolution to definitely resolve filamentary
or spiral structure in the other objects in our survey.

\subsection{Surface Brightness Relations}
\label{surfacebrightness}

Secure identification of binarity is 
critical -- \citet{monnier99} emphasize that if
all dusty WR stars are in binaries that this has profound implications
on massive stellar evolution and the nature of carbon-rich (WC) WR
stars in particular.  One method to achieve this end is to see if all
the objects in our survey follow similar {\em surface brightness
  relations}.  In order to calculate a specific surface brightness
$S_\lambda$ we applied the following formula:

\begin{equation}
S_\lambda = \frac{ F_\lambda 10^{-({\rm mag}/2.5)} } {0.68 \pi \Theta_{\rm FWHM}^2 }
\end{equation}
where $F_\lambda$ represent the flux density for a zero
magnitude star (here using units of W m$^{-2} \mu$m$^{-1}$), mag is the
magnitude of the WR dust shell at the given wavelength (correcting for
stellar emission when appropriate -- see \S\ref{corrections}), and
$\Theta_{\rm FWHM}$ is the FWHM of the Gaussian fit.  Note the $0.68$
in the formula is a correction factor which converts a Gaussian FWHM
size estimate to the equivalent uniform disk for a better definition
of ``Surface Brightness.''  The final units of S$_\lambda $ are W
m$^{-2} \mu$m$^{-1}$ sr$^{-1}$.

\begin{figure}

\begin{center}
\includegraphics[angle=90,width=6in]{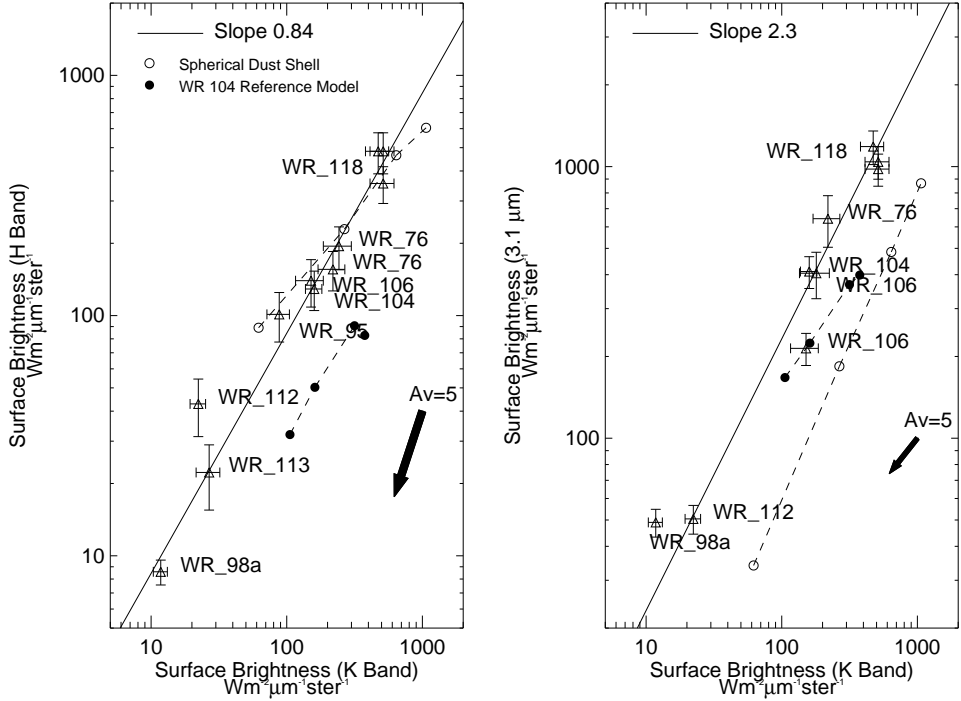}
\caption{Here we show the surface brightness relations for dusty
  Wolf-Rayet systems.  We plot the best-fit linear relation corresponding to
  $T_{\rm color} \sim 1000~K$ (left panel) and $T_{\rm color}\sim
  650~K$ (right panel), assuming optical constants for \citet{zubko98}
  dust\label{fig_surface}.  The open circles show the surface
  brightness relations for a series of spherically-symmetric dust
  shell models with $\tau_{2.2\mu}=$0.01, 0.1, 0.3, 0.67 (higher $\tau$ yields
higher $S_\lambda$).  The filled
  circles show the same relations for the WR~104 Reference Model of
  \citet{harries2004} for inclination angles of $i=$0, 30, 60,
  90$\arcdeg$ (higher $i$ yields higher $S_\lambda$).
The arrow shows how the surface brightness changes for interstellar
reddening of $A_V=5$, assuming standard ISM dust.  }
\end{center}
\end{figure}

Figure~\ref{fig_surface} shows the observed surface brightness
relations.  The objects in the survey all appear to  follow the same surface
brightness relations, indicating all the systems share a similar
near-infrared emission mechanism.  Assuming the emission comes from
optically-thin carbon dust \citep[described by the optical constants
of ][]{zubko98}, we can estimate the color temperature, and we find
$T_{\rm color} \sim 1000~K$ (considering the relation between
1.65$\mu$m and 2.2$\mu$m) and $T_{\rm color}\sim 650~K$ (considering
2.2$\mu$m and 3.08$\mu$m).  The fact that the surface brightness
relations can not be described by a single temperature reflects the
fact that the dust from a wide range of temperatures contributes to the
near-IR emission.

We compared our observed surface brightness relations with the
predicted relations from two models. Firstly, we used the radiative
transfer code DUSTY \citep{dusty} to fit spherically-symmetric dust
shell models \citep[assuming uniform outflow
and dust constants of][]{zubko98}  to the SED.  We could vary the amount of 
dust optical
depth and calculate the model surface brightness subject to an overall
fit to the SED.  We present the calculated surface brightness
relations in Figure~\ref{fig_surface}, spanning
$\tau_{2.2\mu}=$0.01--0.77.  We note that the SED fits were not perfect
(usually under-predicting 3.1$\mu$m fluxes), and thus we take these
relations as representative but not fully optimized.

\citet{harries2004} recently calculated the emergent flux from a
series of spiral dust shell models using the 3-D (Monte Carlo)
radiative transfer code TORUS \citep{harries2000}.  We refer to this as
the ``WR 104 Reference Model'' and used this sophisticated physical
model to calculate the surface brightness relations as a function of
inclination angle (the model only reasonably fits the WR~104 SED for
low inclinations).  These results can also be found on
Figure~\ref{fig_surface}.

Comparing the model calculations to the data, we find that neither of
these models are a good fit to all the 1.65$\mu$m, 2.2$\mu$m, and
3.1$\mu$m surface brightness relations, although the models can
reproduce some of the observations.  More generically, this figure
shows us that any particular dust shell geometry yields a family of
surface brightness relations.  The tightly-correlated surface
brightness relations measured for our sample is suggestive of a shared
and distinctive dust shell geometry, although neither the
spherically-symmetric nor 3-D WR 104 Reference Model can fully explain
our results.  One could improve the fits of a given model to the data
by tuning the adopted optical constants (which are not well known),
although the differences {\em between} models would persist.  Before
discussing the implications of this, we look for more information in
the ``size ratio'' relations.

\subsection{Size Ratio Relations}
\label{sizes}
\begin{figure}

\begin{center}
\includegraphics[angle=90,width=6in]{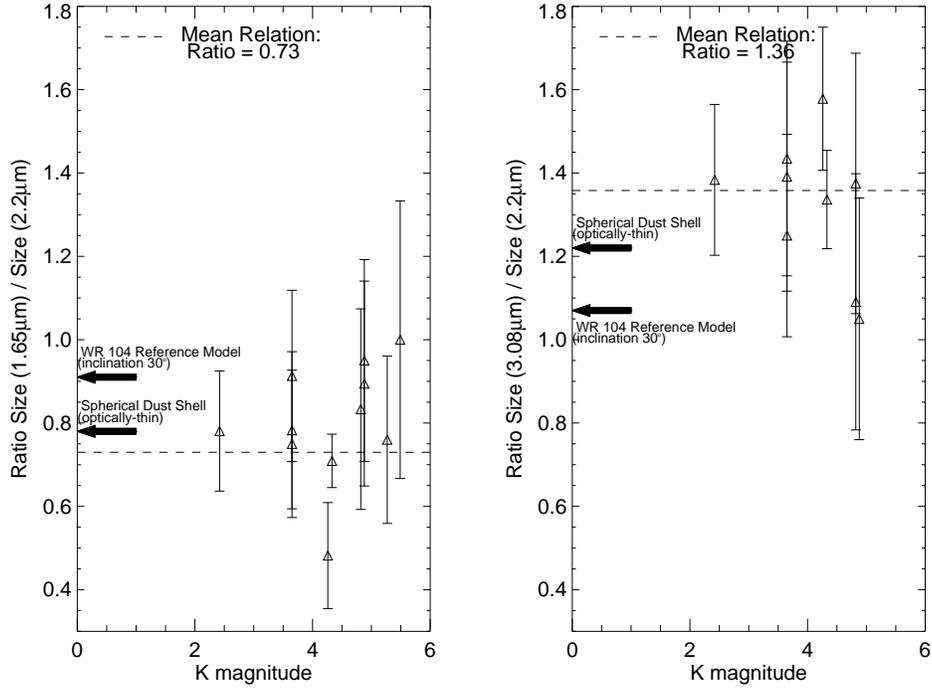}
\caption{Here we plot near-infrared characteristic size ratios vs
K-band magnitude.  The left panel shows the results for
$\frac{ {\rm FWHM~at~} 1.65\mu m}{{\rm FWHM~at~} 2.2\mu m}$, while the right panel shows
the results for $\frac{ {\rm FWHM~at~} 3.08\mu m}{{\rm FWHM~at~} 2.2\mu m}$. 
We plot (dashed line) the mean size ratio and also mark with arrows the
expected size ratios for (a) optically-thin, spherically-symmetric dust shell and (b) the WR~104 ``Reference Model'' of \citet{harries2004} viewed from 30$\arcdeg$ inclination angle.\label{fig_ratios}
} 
6\end{center}
\end{figure}

Another possibly distinctive (and distance-independent) 
signature to differentiate spherically-symmetric
dust shells from colliding wind systems is the ratio of dust
shell characteristic sizes at different wavelengths.  We have
collected all of the size ratios and presented these (with errors) in
Figure~\ref{fig_ratios}.  Assuming all the objects are drawn from the
same class, we find the following {\em mean} size relations:
\begin{eqnarray}
R_{1.65\mu m / 2.2\mu m} = \frac{ {\rm FWHM~at~} 1.65\mu m}{{\rm FWHM~at~} 2.2\mu m} & = & 0.73 \pm 0.05 \\
R_{3.08\mu m / 2.2\mu m} = \frac{ {\rm FWHM~at~} 3.08\mu m}{{\rm FWHM~at~} 2.2\mu m} & = & 1.36 \pm 0.07
\end{eqnarray}

This compares favorably with calculations for an optically-thin,
spherically-symmetric dust shell ($R_{1.65\mu m / 2.2\mu m}=0.78$, $
R_{3.08\mu m / 2.2\mu m}=1.22$), as performed using DUSTY \citep{dusty}
and tuned to fit the SED.  
We have also calculated the size ratios
predicted by the WR~104 reference model discussed in the last section.
As shown in \citet{harries2004}, this model does a remarkable job in
fitting the morphology, SED, and overall size.  The series of dust
shell models used to calculate the size relations (as well as surface
brightness relations) can be found in Figure~\ref{fig_harries}.
Analysis of these synthetic images yields the following mean size
relations:

\begin{eqnarray*}
{\rm inclination~0}\arcdeg:\qquad R_{1.65\mu m / 2.2\mu m} & = & 0.92 \\
 R_{3.08\mu m / 2.2\mu m} & = & 1.04 \\
{\rm inclination~30}\arcdeg:\qquad R_{1.65\mu m / 2.2\mu m} & = & 0.91 \\
 R_{3.08\mu m / 2.2\mu m} & = & 1.07 \\
{\rm inclination~60}\arcdeg:\qquad R_{1.65\mu m / 2.2\mu m} & = & 0.95 \\
 R_{3.08\mu m / 2.2\mu m} & = & 1.14 \\
{\rm inclination~90}\arcdeg:\qquad R_{1.65\mu m / 2.2\mu m} & = & 0.98 \\
 R_{3.08\mu m / 2.2\mu m} & = & 1.28 
\end{eqnarray*}

Paradoxically, the simple spherically-symmetric model fits the WR~104 size
relations {\em better} than the model made specifically for WR~104.
This can be explained if the optical depth of the
shell in the reference model of \citet{harries2004} is somewhat too
high.  High $\tau$ in the inner windings of the spiral casts a strong
shadow across the nebula which cause the temperature profile to be
very steep (analogous to the temperature profile for an
optically-thick dust shell).  Most likely the observed size relations
can be accommodated by the pinwheel model by adjusting the
distribution of dust in the inner most part of the nebulae.

\subsection{Comments on individual objects}

\begin{figure}

\begin{center}
\includegraphics[angle=90,width=5.5in]{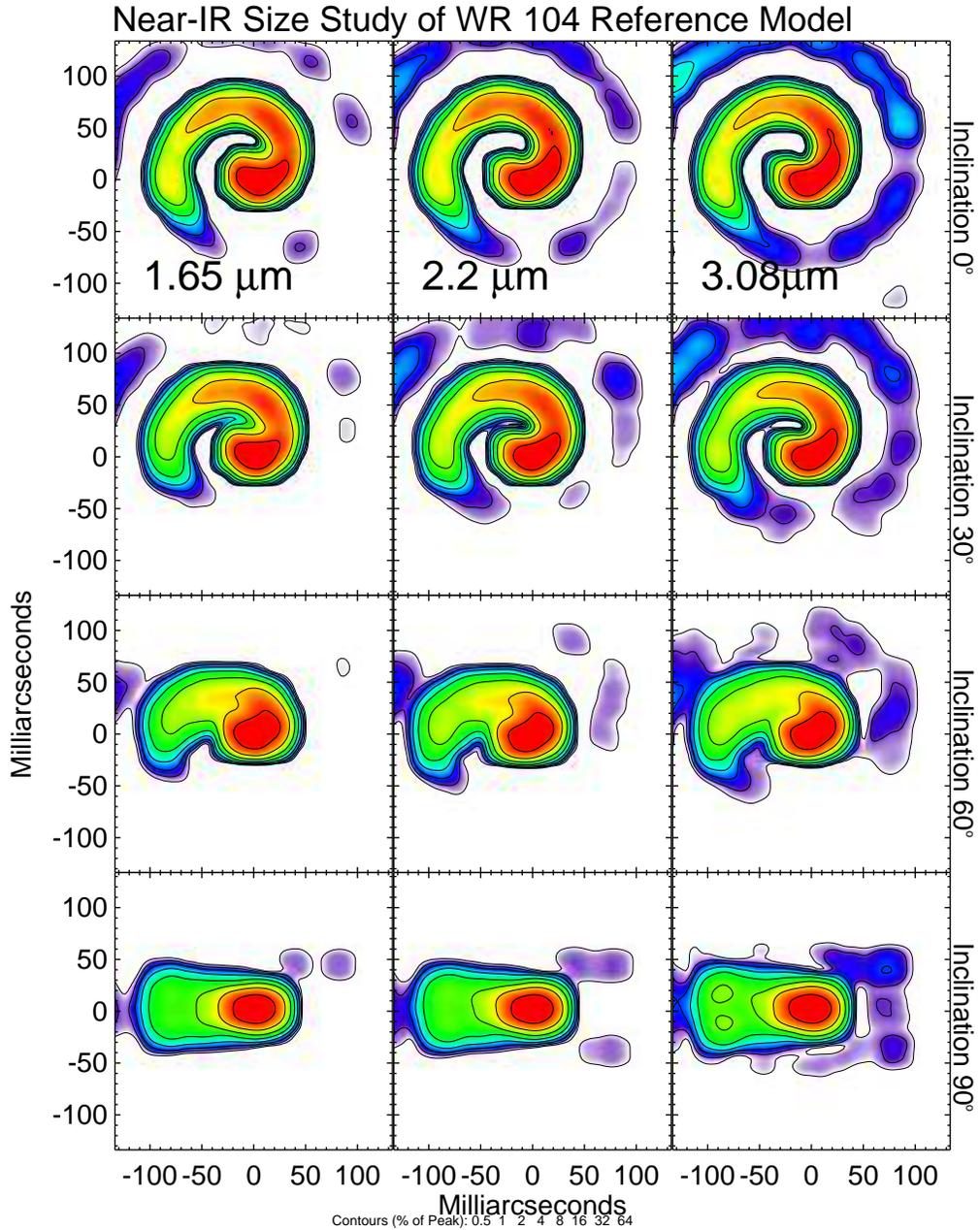}
\caption{Near-IR Size Study of WR 104 Reference Model. This figure shows a gallery of synthetic images
of the WR~104 reference model described in \citet{harries2004},
smoothed to a uniform angular resolution of 20~mas.
These images were used to derive the surface brightness and size ratio comparisons 
described in \S\ref{surfacebrightness} \& \ref{sizes}.  The 3 columns correspond to 
effective wavelengths of 1.65$\mu$m, 2.2$\mu$m, and 3.08$\mu$m, 
while the 4 rows correspond to viewing angles of inclination
0$\arcdeg$, 30$\arcdeg$, 60$\arcdeg$, and 90$\arcdeg$.\label{fig_harries}
} 
\end{center}
\end{figure}

{\bf WR~112:} We know that WR~112 is in a colliding wind binary system
(like WR~98a and WR~104) from its clear non-thermal radio emission
\citep{chapman99, monnier2002c}.  Furthermore, we believe that we are
viewing the binary near the orbital plane (not face-on) because the
non-thermal radio emission is highly variable, suggesting that our
line-of-sight periodically passes through the optically-thin O-star
wind \citep[as is well-documented for WR~140,][]{wb95}.  In this
context, it is not surprising that the colliding wind dust spiral
takes on a more complicated shape due to projection effects \citep[see
discussion of this in][]{monnier2002b,harries2004}.  Indeed, the
$i=$60$\arcdeg$ model for WR~104 (Figure~\ref{fig_harries}) bears an
obvious similarity to the ``horseshoe'' structure that we observed 
for WR~112 (middle-panel of Figure~\ref{fig112}).

This suggestion of an edge-one viewing angle contradicts the
apparent (near) face-o dust spiral geometry seen in
the mid-IR \citep{marchenko2002}.  We speculate that projection effects of the
outer spiral windings (seen at high inclination) could also explain the mid-IR data, 
yielding bright arcs and filaments that 
may resemble a face-on spiral outflow \citep[e.g.,][]{monnier2002c}. Alternatively,
if the underlying colliding wind system is indeed viewed from a face-on position as implied
by the mid-IR images, then the variable radio emission might be caused by a, as yet undetected,
third stellar component to the system.  
Further analysis of this system will be included in a future
paper where we will present multi-epoch near- and mid-IR images as well as VLA
monitoring photometry.

{\bf WR~113 CV Ser:} \citet{vdh2001} lists WR~113 as a nearly edge-on
binary system (inclination$\sim$70$\arcdeg$) with colliding winds.  
We expect such edge-on systems to
show significant dust shell elongations in near-IR
(see Figure~\ref{fig_harries}), although our data indicates the dust emission
is symmetric (Ratio $r=\rm{FWHM}_{\rm{minor}} / \rm{FWHM}_{\rm{major}} \simge 0.8$).
This higher level of symmetry could result if the cone opening angle of
colliding wind interface was larger than found for WR~104 or if there was 
significant dust entrained in the WR outflow in a more spherical pattern.
Perhaps the hint of a 3rd component in the system \citep{niemela1996} could
also explain this surprising lack of dust shell asymmetry in this object.

{\bf WR~118:} Our K-band size estimate is compatible with the earlier 
speckle measurement of \citet{yudin2001}.    

{\bf Quintuplet Stars:} \citet{quint2006} recently reported resolving
two bright Quintuplet WC stars into pinwheel nebulae.  
These objects
show a size ratio $R_{3.08\mu m / 2.2\mu m}\sim 2$, larger than for the stars
reported here, along with a correspondingly low 3.1$\mu$m surface
brightness.  We have not included these stars in this paper given the
difficulty in correcting for extinction and the possible importance of
local heating in this unusually dense and active cluster.

\section{Conclusions and Future Work}

We have presented a multi-wavelength survey of near-infrared angular
sizes of dusty Wolf-Rayet systems from the Keck Aperture Masking
program.  Aperture synthesis images were presented for WR~112 for the
first time and we found strong evidence for interacting/colliding
winds.  In addition, we presented new epochs of WR~104 and
WR~98a images as well as the first results at 3.1$\mu$m, confirming
the spiral nature of the dust shells for WR~104 and WR~98.

Using these data, we discovered tightly-correlated surface brightness
relations and also common size ratios between different near-infrared
bands.  The observed relations could not be reproduced in detail using
either a spherically-symmetric dust shell model or the 3-D radiative
transfer model of WR~104 \citep{harries2004}. The high-quality data
presented here will act as an observational foundation for a new
generation of modelling efforts.

We find these results to be compelling {\em indirect} evidence that 
all these dusty WR here share a common emission geometry, presumably related
to the obvious spiral distribution of WR~104 and WR~98a 
\citep[and confirmed recently for
two Quintuplet WC stars;][]{quint2006}.  
Thus, while not conclusive, this study can be viewed as further evidence that WC Wolf-Rayet stars 
are associated with binarity \citep{monnier99, williams2000}.

In order to make further progress on identifying binarity, we must pursue multiple 
approaches.
While traditional direct spectroscopic identifications of binarity have
proved difficult \citep[e.g.,][]{williams2000}, other promising methods
have emerged for unambiguously determining if dusty WR systems are binaries.
We recommend the following
strategies  be pursued:

\begin{itemize}
\item Higher resolution infrared data.  A factor of 3 longer
  baselines will allow unambiguous detection of asymmetry and non-zero
  closure phases, if spirals and/or filaments exist in our sample.
  This corresponds to about 30~m baselines, easily achievable with the
  current interferometers (the Very Large Telescope is particularly
  well-suited to this).  The main challenge will be the low visible
  flux which makes tip-tilt tracking difficult -- infrared star
  trackers would be helpful in this regard.

\item Visible or infrared photometry. Photometric fluctuations have
  been reported for WR~98a \citep{monnier99} and WR~104
  \citep{kato2002}.  The variations seem to correlate to the known
  orbital periods and we strongly encourage observers to monitor all
  dusty WC stars to establish the periods of the putative underlying
  binaries. This method can work for objects at large distances (e.g.,
  galactic center, Local Group), however one has to contend with
  intrinsic variability and have the determination to monitor sources 
  over many years
  \citep[or decades;][]{williams1990}.

\item Non-thermal radio emission. For systems with periods $\simge$1~year, we expect detectable
  non-thermal radio emission from the colliding winds
  \citep{dougherty2000, monnier2002}.  This requires sensitive
  measurements at multiple frequencies but should be possible and will
  yield unambiguous results.  Recent successful attempts to use this method 
  include \citet{leitherer97}, \citet{chapman99}, \citet{monnier2002c},
and \citet{cappa2004}.  

\end{itemize}

While skeptics will not be convinced yet that the WC phenomena has
a necessary connection to binarity, the evidence in favor of this scenario continues to
accumulate.  The proposed observational efforts to conclusively establish binarity
complement on-going theoretical studies to explore the far-reaching
consequences of the role of binarity in the theory of massive star
evolution.  Indeed, a complete understanding of Gamma Ray Burst
progenitors, the Galactic Black Hole population, and the high-mass end
of the Initial Mass Function  hinges on an accurate picture of
massive binary stellar evolution.


\acknowledgments
This work would not have been possible without 
the support of Dr. Charles Townes (U.C. Berkeley); much of data in this
paper were acquired during observing time granted to him.
This research has made use of the SIMBAD database,
operated at CDS, Strasbourg, France. This publication makes use of
data products from the Two Micron All Sky Survey (2MASS), which is a joint
project of the University of Massachusetts and the Infrared Processing
and Analysis Center/California Institute of Technology, funded by the
National Aeronautics and Space Administration and the National Science
Foundation. 
The data presented herein were obtained at
the W.M. Keck Observatory, which is operated as a scientific
partnership among the California Institute of Technology, the
University of California and the National Aeronautics and Space
Administration.  The Keck Observatory was made possible by the
generous financial support of the W.M. Keck Foundation.  
The authors wish to recognize and acknowledge the very significant
cultural role and reverence that the summit of Mauna Kea has always
had within the indigenous Hawaiian community.  We are most fortunate
to have the opportunity to conduct observations from this mountain.

\bibliographystyle{apj}
\bibliography{apj-jour,WRSizes,HerbigSizes,Thesis,IONIC3,WR140,Review,iKeck,RX_Boo}

\begin{thebibliography}{53}
\expandafter\ifx\csname natexlab\endcsname\relax\def\natexlab#1{#1}\fi

\bibitem[{{Cappa} {et~al.}(2004){Cappa}, {Goss}, \& {van der
  Hucht}}]{cappa2004}
{Cappa}, C., {Goss}, W.~M., \& {van der Hucht}, K.~A. 2004, \aj, 127, 2885

\bibitem[{{Chapman} {et~al.}(1999){Chapman}, {Leitherer}, {Koribalski},
  {Bouter}, \& {Storey}}]{chapman99}
{Chapman}, J.~M., {Leitherer}, C., {Koribalski}, B.~., {Bouter}, R., \&
  {Storey}, M. 1999, \apj, 518, 890

\bibitem[{{Cherchneff} {et~al.}(2000){Cherchneff}, {Le Teuff}, {Williams}, \&
  {Tielens}}]{cherchneff2002}
{Cherchneff}, I., {Le Teuff}, Y.~H., {Williams}, P.~M., \& {Tielens},
  A.~G.~G.~M. 2000, \aap, 357, 572

\bibitem[{{Cutri} {et~al.}(2003){Cutri}, {Skrutskie}, {van Dyk}, {Beichman},
  {Carpenter}, {Chester}, {Cambresy}, {Evans}, {Fowler}, {Gizis}, {Howard},
  {Huchra}, {Jarrett}, {Kopan}, {Kirkpatrick}, {Light}, {Marsh}, {McCallon},
  {Schneider}, {Stiening}, {Sykes}, {Weinberg}, {Wheaton}, {Wheelock}, \&
  {Zacarias}}]{cutri2003}
{Cutri}, R.~M., {Skrutskie}, M.~F., {van Dyk}, S., {Beichman}, C.~A.,
  {Carpenter}, J.~M., {Chester}, T., {Cambresy}, L., {Evans}, T., {Fowler}, J.,
  {Gizis}, J., {Howard}, E., {Huchra}, J., {Jarrett}, T., {Kopan}, E.~L.,
  {Kirkpatrick}, J.~D., {Light}, R.~M., {Marsh}, K.~A., {McCallon}, H.,
  {Schneider}, S., {Stiening}, R., {Sykes}, M., {Weinberg}, M., {Wheaton},
  W.~A., {Wheelock}, S., \& {Zacarias}, N. 2003, VizieR Online Data Catalog,
  2246, 0

\bibitem[{{Danchi} {et~al.}(2001){Danchi}, {Tuthill}, \&
  {Monnier}}]{danchi2001}
{Danchi}, W.~C., {Tuthill}, P.~G., \& {Monnier}, J.~D. 2001, \apj, 562, 440

\bibitem[{{de Marco} \& {Schmutz}(1999)}]{demarco1999}
{de Marco}, O. \& {Schmutz}, W. 1999, \aap, 345, 163

\bibitem[{{De Marco} {et~al.}(2000){De Marco}, {Schmutz}, {Crowther},
  {Hillier}, {Dessart}, {de Koter}, \& {Schweickhardt}}]{demarco2000}
{De Marco}, O., {Schmutz}, W., {Crowther}, P.~A., {Hillier}, D.~J., {Dessart},
  L., {de Koter}, A., \& {Schweickhardt}, J. 2000, \aap, 358, 187

\bibitem[{{Dougherty} {et~al.}(2005){Dougherty}, {Beasley}, {Claussen},
  {Zauderer}, \& {Bolingbroke}}]{dougherty2005}
{Dougherty}, S.~M., {Beasley}, A.~J., {Claussen}, M.~J., {Zauderer}, B.~A., \&
  {Bolingbroke}, N.~J. 2005, \apj, 623, 447

\bibitem[{{Dougherty} \& {Williams}(2000{\natexlab{a}})}]{dw2000}
{Dougherty}, S.~M. \& {Williams}, P.~M. 2000{\natexlab{a}}, \mnras, 319, 1005

\bibitem[{{Dougherty} \& {Williams}(2000{\natexlab{b}})}]{dougherty2000}
---. 2000{\natexlab{b}}, \mnras, 319, 1005

\bibitem[{Efron \& Tibshirani(1993)}]{bootstrap}
Efron, B. \& Tibshirani, R.~J. 1993, An introduction to the bootstrap ({New
  York}: Chapman and Hall)

\bibitem[{{ESA}(1997)}]{hipparcos}
{ESA}. 1997, VizieR Online Data Catalog, 1239, 0

\bibitem[{{Gull} \& {Skilling}(1983)}]{mem83}
{Gull}, S.~F. \& {Skilling}, J. 1983, in Indirect Imaging. Measurement and
  Processing for Indirect Imaging. Proceedings of an International Symposium
  held in Sydney, Australia, August 30-September 2, 1983. Editor, J.A. Roberts;
  Publisher, Cambridge University Press, Cambridge, England, New York, NY,
  1984. LC \# QB51.3.E43 I53 1984. ISBN \# 0-521-26282-8. P.267, 1983, 267+

\bibitem[{{Harries}(2000)}]{harries2000}
{Harries}, T.~J. 2000, \mnras, 315, 722

\bibitem[{{Harries} {et~al.}(2004){Harries}, {Monnier}, {Symington}, \&
  {Kurosawa}}]{harries2004}
{Harries}, T.~J., {Monnier}, J.~D., {Symington}, N.~H., \& {Kurosawa}, R. 2004,
  \mnras, 350, 565

\bibitem[{Harrison \& Goodrich(1999)}]{nircmanual}
Harrison, W. \& Goodrich, R.~W. 1999, The {NIRC} User's Manual, maintained by
  the {Keck} Observatory

\bibitem[{{Ivezic} {et~al.}(1999){Ivezic}, {Nenkova}, \& {Elitzur}}]{dusty}
{Ivezic}, Z., {Nenkova}, M., \& {Elitzur}, M. 1999, User Manual for DUSTY,
  Tech. rep., University of Kentucky

\bibitem[{{Kato} {et~al.}(2002){Kato}, {Haseda}, {Yamaoka}, \&
  {Takamizawa}}]{kato2002}
{Kato}, T., {Haseda}, K., {Yamaoka}, H., \& {Takamizawa}, K. 2002, \pasj, 54,
  L51

\bibitem[{{Leitherer} {et~al.}(1997){Leitherer}, {Chapman}, \&
  {Koribalski}}]{leitherer97}
{Leitherer}, C., {Chapman}, J.~M., \& {Koribalski}, B. 1997, \apj, 481, 898

\bibitem[{{Marchenko} {et~al.}(2002){Marchenko}, {Moffat}, {Vacca},
  {C{\^o}t{\'e}}, \& {Doyon}}]{marchenko2002}
{Marchenko}, S.~V., {Moffat}, A.~F.~J., {Vacca}, W.~D., {C{\^o}t{\'e}}, S., \&
  {Doyon}, R. 2002, \apjl, 565, L59

\bibitem[{{Matthews} {et~al.}(1996){Matthews}, {Ghez}, {Weinberger}, \&
  {Neugebauer}}]{matthews96}
{Matthews}, K., {Ghez}, A.~M., {Weinberger}, A.~J., \& {Neugebauer}, G. 1996,
  \pasp, 108, 615+

\bibitem[{{Millan-Gabet} {et~al.}(2001){Millan-Gabet}, {Schloerb}, \&
  {Traub}}]{rmg2001}
{Millan-Gabet}, R., {Schloerb}, F.~P., \& {Traub}, W.~A. 2001, \apj, 546, 358

\bibitem[{{Moffat} {et~al.}(1987){Moffat}, {Lamontagne}, {Williams}, {Horn}, \&
  {Seggewiss}}]{moffat87}
{Moffat}, A. F.~J., {Lamontagne}, R., {Williams}, P.~M., {Horn}, J., \&
  {Seggewiss}, W. 1987, \apj, 312, 807

\bibitem[{{Monnier} {et~al.}(2002{\natexlab{a}}){Monnier}, {Greenhill},
  {Tuthill}, \& {Danchi}}]{monnier2002winds}
{Monnier}, J.~D., {Greenhill}, L.~J., {Tuthill}, P.~G., \& {Danchi}, W.~C.
  2002{\natexlab{a}}, in ASP Conf. Ser. 260: Interacting Winds from Massive
  Stars, ed. A.~F.~J. {Moffat} \& N.~{St-Louis}, 331--+

\bibitem[{{Monnier} {et~al.}(2002{\natexlab{b}}){Monnier}, {Greenhill},
  {Tuthill}, \& {Danchi}}]{monnier2002c}
{Monnier}, J.~D., {Greenhill}, L.~J., {Tuthill}, P.~G., \& {Danchi}, W.~C.
  2002{\natexlab{b}}, \apj, 566, 399

\bibitem[{{Monnier} {et~al.}(2002{\natexlab{c}}){Monnier}, {Greenhill},
  {Tuthill}, \& {Danchi}}]{monnier2002}
---. 2002{\natexlab{c}}, \apj, 566, 399

\bibitem[{{Monnier} {et~al.}(2005){Monnier}, {Millan-Gabet}, {Billmeier},
  {Akeson}, {Wallace}, {Berger}, {Calvet}, {D'Alessio}, {Danchi}, {Hartmann},
  {Hillenbrand}, {Kuchner}, {Rajagopal}, {Traub}, {Tuthill}, {Boden}, {Booth},
  {Colavita}, {Gathright}, {Hrynevych}, {Le Mignant}, {Ligon}, {Neyman},
  {Swain}, {Thompson}, {Vasisht}, {Wizinowich}, {Beichman}, {Beletic},
  {Creech-Eakman}, {Koresko}, {Sargent}, {Shao}, \& {van Belle}}]{monnier2005}
{Monnier}, J.~D., {Millan-Gabet}, R., {Billmeier}, R., {Akeson}, R.~L.,
  {Wallace}, D., {Berger}, J.-P., {Calvet}, N., {D'Alessio}, P., {Danchi},
  W.~C., {Hartmann}, L., {Hillenbrand}, L.~A., {Kuchner}, M., {Rajagopal}, J.,
  {Traub}, W.~A., {Tuthill}, P.~G., {Boden}, A., {Booth}, A., {Colavita}, M.,
  {Gathright}, J., {Hrynevych}, M., {Le Mignant}, D., {Ligon}, R., {Neyman},
  C., {Swain}, M., {Thompson}, R., {Vasisht}, G., {Wizinowich}, P., {Beichman},
  C., {Beletic}, J., {Creech-Eakman}, M., {Koresko}, C., {Sargent}, A., {Shao},
  M., \& {van Belle}, G. 2005, \apj, 624, 832

\bibitem[{{Monnier} {et~al.}(2004){Monnier}, {Millan-Gabet}, {Tuthill},
  {Traub}, {Carleton}, {Coud{\'e} du Foresto}, {Danchi}, {Lacasse}, {Morel},
  {Perrin}, {Porro}, {Schloerb}, \& {Townes}}]{monnier2004a}
{Monnier}, J.~D., {Millan-Gabet}, R., {Tuthill}, P.~G., {Traub}, W.~A.,
  {Carleton}, N.~P., {Coud{\'e} du Foresto}, V., {Danchi}, W.~C., {Lacasse},
  M.~G., {Morel}, S., {Perrin}, G., {Porro}, I.~L., {Schloerb}, F.~P., \&
  {Townes}, C.~H. 2004, \apj, 605, 436

\bibitem[{{Monnier} {et~al.}(1999{\natexlab{a}}){Monnier}, {Tuthill}, \&
  {Danchi}}]{monnier99}
{Monnier}, J.~D., {Tuthill}, P.~G., \& {Danchi}, W.~C. 1999{\natexlab{a}},
  \apjl, 525, L97

\bibitem[{{Monnier} {et~al.}(1999{\natexlab{b}}){Monnier}, {Tuthill}, \&
  {Danchi}}]{monnier1999}
---. 1999{\natexlab{b}}, \apjl, 525, L97

\bibitem[{{Monnier} {et~al.}(2002{\natexlab{d}}){Monnier}, {Tuthill}, \&
  {Danchi}}]{monnier2002b}
---. 2002{\natexlab{d}}, \apjl, 567, L137

\bibitem[{{Monnier} {et~al.}(1999{\natexlab{c}}){Monnier}, {Tuthill}, {Lopez},
  {Cruzalebes}, {Danchi}, \& {Haniff}}]{monnier99a}
{Monnier}, J.~D., {Tuthill}, P.~G., {Lopez}, B., {Cruzalebes}, P., {Danchi},
  W.~C., \& {Haniff}, C.~A. 1999{\natexlab{c}}, \apj, 512, 351

\bibitem[{{Niemela} {et~al.}(1996){Niemela}, {Morrell}, {Barba}, \&
  {Bosch}}]{niemela1996}
{Niemela}, V.~S., {Morrell}, N.~I., {Barba}, R.~H., \& {Bosch}, G.~L. 1996, in
  Revista Mexicana de Astronomia y Astrofisica Conference Series, ed.
  V.~{Niemala} \& N.~{Morrell}, 100--+

\bibitem[{{Pauls} {et~al.}(2005){Pauls}, {Young}, {Cotton}, \&
  {Monnier}}]{pauls2005}
{Pauls}, T.~A., {Young}, J.~S., {Cotton}, W.~D., \& {Monnier}, J.~D. 2005,
  \pasp, 117, 1255

\bibitem[{{Ragland} \& {Richichi}(1999)}]{rr99}
{Ragland}, S. \& {Richichi}, A. 1999, \mnras, 302, L13

\bibitem[{{Rajagopal} {et~al.}(2004){Rajagopal}, {Barry}, {Lopez}, {Danchi},
  {Monnier}, {Tuthill}, \& {Townes}}]{rajagopal2004}
{Rajagopal}, J.~K., {Barry}, R., {Lopez}, B., {Danchi}, W.~C., {Monnier},
  J.~D., {Tuthill}, P.~G., \& {Townes}, C.~H. 2004, in New Frontiers in Stellar
  Interferometry, Proceedings of SPIE Volume 5491. Edited by Wesley A. Traub.
  Bellingham, WA: The International Society for Optical Engineering, 2004.,
  p.1120, ed. W.~A. {Traub}, 1120--+

\bibitem[{{Sivia}(1987)}]{sivia87}
{Sivia}, D. 1987, PhD thesis, Cambridge University

\bibitem[{{Skilling} \& {Bryan}(1984)}]{sb84}
{Skilling}, J. \& {Bryan}, R.~K. 1984, \mnras, 211, 111+

\bibitem[{{Tuthill} {et~al.}(2006){Tuthill}, {Monnier}, {Tanner}, {Figer}, \&
  {Ghez}}]{quint2006}
{Tuthill}, P., {Monnier}, J., {Tanner}, A., {Figer}, D., \& {Ghez}, A. 2006,
  Science, in press

\bibitem[{{Tuthill} {et~al.}(1999{\natexlab{a}}){Tuthill}, {Monnier}, \&
  {Danchi}}]{tuthill1999}
{Tuthill}, P.~G., {Monnier}, J.~D., \& {Danchi}, W.~C. 1999{\natexlab{a}},
  \nat, 398, 487

\bibitem[{{Tuthill} {et~al.}(1999{\natexlab{b}}){Tuthill}, {Monnier}, \&
  {Danchi}}]{tuthill99}
---. 1999{\natexlab{b}}, \nat, 398, 487

\bibitem[{{Tuthill} {et~al.}(2000{\natexlab{a}}){Tuthill}, {Monnier}, {Danchi},
  \& {Lopez}}]{tuthill2000b}
{Tuthill}, P.~G., {Monnier}, J.~D., {Danchi}, W.~C., \& {Lopez}, B.
  2000{\natexlab{a}}, \apj, in press

\bibitem[{{Tuthill} {et~al.}(2000{\natexlab{b}}){Tuthill}, {Monnier}, {Danchi},
  {Wishnow}, \& {Haniff}}]{tuthill2000a}
{Tuthill}, P.~G., {Monnier}, J.~D., {Danchi}, W.~C., {Wishnow}, E., \&
  {Haniff}, C.~A. 2000{\natexlab{b}}, \pasp, in press

\bibitem[{{Tuthill} {et~al.}(2000{\natexlab{c}}){Tuthill}, {Monnier}, {Danchi},
  {Wishnow}, \& {Haniff}}]{tuthill2000}
{Tuthill}, P.~G., {Monnier}, J.~D., {Danchi}, W.~C., {Wishnow}, E.~H., \&
  {Haniff}, C.~A. 2000{\natexlab{c}}, \pasp, 112, 555

\bibitem[{{Usov}(1991)}]{usov91}
{Usov}, V.~V. 1991, \mnras, 252, 49

\bibitem[{{van der Hucht}(2001)}]{vdh2001}
{van der Hucht}, K.~A. 2001, New Astronomy Review, 45, 135

\bibitem[{{White} \& {Becker}(1995)}]{wb95}
{White}, R.~L. \& {Becker}, R.~H. 1995, \apj, 451, 352

\bibitem[{{Williams} \& {van der Hucht}(2000)}]{williams2000}
{Williams}, P.~M. \& {van der Hucht}, K.~A. 2000, \mnras, 314, 23

\bibitem[{{Williams} {et~al.}(1990{\natexlab{a}}){Williams}, {van der Hucht},
  {Pollock}, {Florkowski}, {van der Woerd}, \& {Wamsteker}}]{williams1990}
{Williams}, P.~M., {van der Hucht}, K.~A., {Pollock}, A.~M.~T., {Florkowski},
  D.~R., {van der Woerd}, H., \& {Wamsteker}, W.~M. 1990{\natexlab{a}}, \mnras,
  243, 662

\bibitem[{{Williams} {et~al.}(1990{\natexlab{b}}){Williams}, {van der Hucht},
  {Pollock}, {Florkowski}, {van der Woerd}, \& {Wamsteker}}]{williams90}
{Williams}, P.~M., {van der Hucht}, K.~A., {Pollock}, A. M.~T., {Florkowski},
  D.~R., {van der Woerd}, H., \& {Wamsteker}, W.~M. 1990{\natexlab{b}}, \mnras,
  243, 662

\bibitem[{{Williams} {et~al.}(1987){Williams}, {van der Hucht}, \&
  {The}}]{williams1987}
{Williams}, P.~M., {van der Hucht}, K.~A., \& {The}, P.~S. 1987, \aap, 182, 91

\bibitem[{{Yudin} {et~al.}(2001){Yudin}, {Balega}, {Bl{\"o}cker}, {Hofmann},
  {Schertl}, \& {Weigelt}}]{yudin2001}
{Yudin}, B., {Balega}, Y., {Bl{\"o}cker}, T., {Hofmann}, K.-H., {Schertl}, D.,
  \& {Weigelt}, G. 2001, \aap, 379, 229

\bibitem[{{Zubko}(1998)}]{zubko98}
{Zubko}, V.~G. 1998, \mnras, 295, 109+

\end{thebibliography}

\begin{deluxetable}{cccc}
\tablecaption{Properties of NIRC Camera Infrared Filters 
\label{filters}}
\tablewidth{0pt}
\tablehead{ \colhead{Name} & \colhead{Center Wavelength} & \colhead{Bandpass FWHM} & \colhead{Fractional}\\
&\colhead{$\lambda_0$ (\micronn)} & \colhead{$\Delta\lambda$ (\micronn)} &\colhead{ Bandwidth } }
\startdata
FeII & 1.6471 & 0.0176 & 1.1\% \\
H & 1.6575 & 0.333 & 20\% \\
K & 2.2135 & 0.427 & 19\% \\
Kcont & 2.25965 & 0.0531 & 2.3\% \\
CH4 & 2.269 & 0.155 & 6.8\% \\
PAHcs & 3.0825 & 0.1007 &3.3\% 
\enddata
\tablecomments{Reference: The NIRC Manual \citep{nircmanual}}
\end{deluxetable}

\begin{deluxetable}{lllllllllll}
\rotate
\tabletypesize{\tiny}
\tablecaption{Basic Properties of Targets\label{targets}}
\tablewidth{0pt}
\tablehead{
\colhead{Source} & \colhead{RA (J2000)} & \colhead{Dec (J2000)} &
\colhead{V} & \colhead{J} & \colhead{H} & \colhead{K} &
\colhead{Spectral} & \colhead{Distance} &\colhead{Luminosity} & \colhead{Photometry}\\
\colhead{Names} & & &\colhead{mag\tablenotemark{a}} & \colhead{mag\tablenotemark{a}} &\colhead{mag\tablenotemark{a}} &\colhead{mag\tablenotemark{a}} & \colhead{Type}  & \colhead{(kpc)} & \colhead{Log (L/L$_\odot$)\tablenotemark{b}} & \colhead{References}
}
\startdata
$\gamma^2$~Vel, WR~11 & 08 09 31.96 & -47 20 11.8 &  1.81 & 2.15 & 2.25 & 2.10 & WC8 + O7.5III-V (1) & 0.26 (2) & 5.5\tablenotemark{c} (3,4) & 5  \\
WR~48a & 13 12 39.65 & -62 42 55.8 &   &  8.74 & 6.80 & 5.09 & WC8ed + ? (1) & 1.2 (1) &  & 5,9 \\
WR~76  & 16 40 05.3 & -45 41 10 & 15.36 & 8.46 & 6.51 & 4.88 & WC9d (1) &  5.35 (1) &  & 5,9 \\
WR~95  & 17 36 19.76 & -33 26 10.9 & 14.00 & 8.29 & 6.67 & 5.27 & WC9d (1) & 2.09 (1) &  & 5,9 \\
WR~98a & 17 41 13.051 & -30 32 30.34 & & 9.14 & 6.51 & 4.33 & WC8-9vd + OB? (1) & 1.9 (8) & & 9 \\
WR~104 & 18 02 04.123 & -23 37 42.24 & 13.54 & 6.67 & 4.34 & 2.42 & WC9d + B0.5V (1) & 2.3 (6) & 5.4\tablenotemark{c} (7)
& 5 \\
WR~106 & 18 04 43.66 & -21 09 30.7 & 11.93 & 7.94 & 6.28 & 4.82 & WC9d (1) & 2.3 (1) & & 5,9 \\
WR~112 & 18 16 33.489& -18 58 42.47 & 17.7 & 8.68& 6.26 & 4.26 & WC9d +OB? (1) & 4.15 (1) & & 5,9 \\
WR~113 & 18 19 07.36 & -11 37 59.2 & 9.43 &  7.02 & 6.28 & 5.49 & WC8d +O8-9 (1) & 1.79 (1) & & 5,9 \\
WR~118 & 18 31 42.3  & -09 59 15   &  22   & 8.10 & 5.41 & 3.65 & WC9 (1) & 3.13 (1) & & 5,9,10 \\
WR~140 & 20 20 27.98 & +43 51 16.3 & 6.9 & 5.55 & 5.43 & 5.04 & WC7pd + O4-5 (1) & 1.85 (11) & 6.1\tablenotemark{c} (12) & -- \\
\enddata
\tablenotetext{a}{~ These
  magnitudes (V band from Simbad, and J,H.K bands from 2MASS) are merely
  representative since some of the targets are variable.}  
\tablenotetext{b}{~ Luminosity here has been corrected for the adopted distance used in this paper; not all objects
have well-established luminosities due to high extinction.}
\tablenotetext{c}{~ Binary system total luminosity.}
 \tablecomments{References: (1) \citet{vdh2001}, (2) \citet{hipparcos},
(3) \citet{demarco2000}, (4) \citet{demarco1999}, (5) \citet{williams1987},
(6) \citet{tuthill1999}, (7) \citet{harries2004}, (8) \citet{monnier1999},
(9) 2MASS; \citet{cutri2003}, (10) Simbad, (11) \citet{dougherty2005},
(12) \citet{williams1990}
}

\end{deluxetable}

\clearpage

\begin{deluxetable}{lllll}
\scriptsize
\tabletypesize{\tiny}
\tablecaption{Journal of Observations \label{observations}}
\tablewidth{0pt}
\tablehead{
\colhead{Target} & \colhead{Date}   & \colhead{Filter ($\lambda_o$)}  & \colhead{Aperture}  & \colhead{Calibrator\tablenotemark{a}}\\
 & \colhead{(UT)}   & \colhead{($\mu$m)}   & \colhead{Mask}  & \colhead{Names}
}
\startdata
$\gamma^2$~Vel        & 1999 Feb 05 & CH4 (2.269)  & Golay-21 & HD~68553 (K3, 4.39$\pm$0.44 mas) \\
WR~48a                & 2000 Jan 26 & K (2.2135)   & Annulus-36 & HD~115399 (K5, 1.2$\pm$0.7 mas) \\
WR~76                 & 1998 Jun 05 & H (1.6575)   & Annulus-36  & HD~151834 (G3, 1.2$\pm$0.4 mas) \\
                      &             & K (2.2135)   &            & \\
                      &             & CH4 (2.269)  &            &\\
                      & 1999 Apr 26 & H (1.6575)   &            & HD~153258 (K4, 2.4$\pm$0.9 mas) \\
                      &             & K (2.2135)   &            &\\
                      &             & PAHcs (3.0825) &           &\\
WR~95                 & 2000 Jun 24 & H (1.6575) & Annulus-36& HD~158774 (M0, 2$\pm$2 mas) \\
                      &             & K (2.2135) &           & \\
WR~98a                & 2000 Jun 24 &  H (1.6575) & Annulus-36& HD~158774  \\
                      &             &  K (2.2135) &           & \\
                     &             & CH4 (2.269) &            & Hd~163428 (K5, 4.1$\pm$1.0 mas)\\ 
                     &             & PAHcs (3.0825) &           &\\
WR~104               & 2000 Jun 24 &  H (1.6575) & Annulus-36& HD~165813 (M0, 1.7$\pm$0.9 mas)\\
                    &             & CH4 (2.269)&            & HD 167036 (K2, 3.0$\pm$0.6 mas) \\ 
                     &             &PAHcs (3.0825)  &           &  HD~167036\\
WR~106               & 1998 Jun 05 &  H (1.6575) & Annulus-36& HD~164124 (K2, 1.1$\pm$0.3 mas)\\
                     &             & CH4 (2.269) &            & HD 167036\\
                     &             & PAHcs (3.0825) &           &  HD~167036\\
                     & 1999 Jul 30 & K (2.2135)   &  & \\
                     &             &  PAHcs (3.0825)  &  & \\
WR~112               & 2000 Jun 24 &  H (1.6575) & Annulus-36& HD~165813 \\
                     &             &  K (2.2135)  &           & \\
                     &             & CH4 (2.269)&            & HD~167036\\ 
                     &             &  PAHcs (3.0825) &           &\\
WR~113               & 2000 Jun 24 & H (1.6575) & Annulus-36& HD 168366 (K2, 1.03$\pm$0.12 mas)\\
                     &             & K (2.2135) &           & HD 168366\\
WR~118               & 1998 Jun 05 & H (1.6575) & Annulus-36& HD~168000 (K3, 1.42$\pm$0.8 mas)\\
                     &             & CH4 (2.269) &            & \\
                     &             & PAHcs (3.0825)  &           & HD~175775 (K0, 3.5$\pm$1.1 mas)\\
                     & 1998 Jun 06 &  H (1.6575) & Annulus-36& HD~170474 (K0, 1.3$\pm$0.3 mas)\\
                     &             &CH4 (2.269) &            & HD~173074 (M, 3.4$\pm$1.4 mas)\\
                     &             & PAHcs (3.0825)  &           & HD~175775\\
                     & 1999 Apr 26 &  H (1.6575) & Annulus-36& HD~168000 \\
                     &             & CH4 (2.269) &            & \\
                     &             & PAHcs (3.0825) &           & HD~175775 \\
WR~140               & 1999 Jul 30 & FeII (1.6471) & FFA     & HD~193631 (K0, 0.5$\pm$0.1 mas) \\
                     &             & Kcont (2.25965) & FFA & \\
                     & 2001 Jul 30 & FeII (1.6471) & FFA     & HD~192867 (M1, 1.6$\pm$0.6 mas)  \\
                     &             &  Kcont (2.25965) & FFA & \\
                     &             & PAHcs (3.0825)  & FFA & HD~192867 \\
\enddata
\tablenotetext{a}{Calibrator size estimates made using {\em getCal},
maintained and distributed by the Michelson Science Center (http://msc.caltech.edu).}
\end{deluxetable}

\begin{deluxetable}{llcccl}
\tabletypesize{\scriptsize}
\scriptsize
\tablecaption{Mean Infrared Size Results \label{results1}}
\tablewidth{0pt}
\tablehead{
\colhead{Target} & \colhead{Date}   & \multicolumn{3}{c}{Gaussian FWHM (mas)\tablenotemark{a,b}} & Comments \\
                 & \colhead{(UT)}   & \colhead{1.65$\mu$m} & \colhead{2.2$\mu$m} & \colhead{3.08$\mu$m} &     
}
\startdata
$\gamma^2$~Vel   & 1999 Feb 05 &   --     &   7$\pm$7 & --       &       \\
WR~48a           & 2000 Jan 26 &   --     &  63$\pm$3\tablenotemark{(c)} & --       &  Fit to ``large-scale'' component only.      \\
WR~76            & 1998 Jun 05 & 17$\pm$3 &  19$\pm$4 & --       &       \\
                 & 1999 Apr 26 & 19$\pm$3 &  20$\pm$4 & 21$\pm$4 &       \\
WR~95            & 2000 Jun 24 & 16$\pm$3\tablenotemark{(c)} &  23$\pm$4\tablenotemark{(c)} & --       &       \\
WR~98a           & 2000 Jun 24 & 73$\pm$4\tablenotemark{(c)} & 110$\pm$7 &147$\pm$9 &  Fit to ``large-scale'' component only. \\        
                &             & r$=$0.56 & r$=$0.74  & r$=$0.81 &  Elliptical (PA$\sim$120\arcdegg).\\

WR~104           & 2000 Jun 24 & 57$\pm$9 &  73$\pm$7 & 101$\pm$9 &  Fit to ``large-scale'' component only. \\
                 &             & r$=$0.84 &  r$=$0.88 & r$=$0.85  &  Elliptical (PA$\sim$30\arcdegg).\\
WR~106           & 1998 Jun 05 & 17$\pm$3\tablenotemark{(c)} &  23$\pm$4\tablenotemark{(c)} & 33$\pm$3  & \\
                 & 1999 Jul 30 & --       &  21$\pm$4\tablenotemark{(c)} & 24$\pm$4  & \\
WR~112           & 2000 Jun 24 &40$\pm$10 &  83$\pm$7 & 131$\pm$9  &  Fit to ``large-scale'' component only.\\
                 &             &          &  r$=$0.87 &         &  Elliptical (PA$\sim$130).\\
WR~113           & 2000 Jun 24 & 19$\pm$2\tablenotemark{(c)} &  27$\pm$3\tablenotemark{(c)} & --      & \\
WR~118           & 1998 Jun 05 & 18$\pm$3 &  24$\pm$4 & 30$\pm$3  & \\
                 & 1998 Jun 06 & 18$\pm$3 &  23$\pm$4 & 32$\pm$3  & \\
                 & 1999 Apr 26 & 21$\pm$3 &  23$\pm$4 & 33$\pm$3  & \\
WR~140           & 1999 Jul 30 & $<$15    &     $<$13 &  --       & Pre-outburst \\
                 & 2001 Jul 30 & 268$\pm$17 & 228$\pm$12& 263$\pm$13 & Two-component model: Gaussian $+$ Point \\
                 &             & 39\% dust  & 64\% dust & 78\% dust       & \\
\enddata
\tablenotetext{a}{These characteristic sizes are {\em not} corrected for the
presence of the star (which contribute significant near-IR flux in some cases).}
\tablenotetext{b}{If visibility data are manifestly elliptical, then we report both the
mean FWHM and the Ratio $r=\rm{FWHM}_{\rm{minor}} / \rm{FWHM}_{\rm{major}}$. } 
\tablenotetext{c}{See Table~\ref{results2} for corrected sizes.}
\end{deluxetable}

\begin{deluxetable}{lcccc}
\scriptsize
\tablecaption{Characteristic sizes of Wolf-Rayet dust shells, corrected for stellar contribution
 \label{results2}}
\tablewidth{0pt}
\tablehead{
\colhead{Target}   & \multicolumn{2}{c}{Dust Fraction (\%)} & \multicolumn{2}{c}{Gaussian FWHM (mas)\tablenotemark{a}}  \\
                   & \colhead{1.65$\mu$m} & \colhead{2.2$\mu$m}   & \colhead{1.65$\mu$m} & \colhead{2.2$\mu$m}  
}
\startdata
WR~48a           &  -- & 85\% & --     &  70$\pm$5     \\
WR~95             & 73\% & 89\% & 19$\pm$4 &  25$\pm$4   \\
WR~98a           & 90\% & -- & 78$\pm$5 & -- \\
WR~106            & 78\% & 93\% &  20$\pm$4 &  24$\pm$5  \\
                  & -- & 93\% & --       &  22$\pm$5  \\
WR~113            & 38\% & 65\% & 35$\pm$10 &  35$\pm$6  \\
\enddata
\tablenotetext{a}{These characteristic sizes are corrected for the
presence of the star (which contribute significant near-IR flux in some cases; see
\S\ref{corrections}).  Refer to Table~\ref{results1} for size measurements for rest of
sample which do not require corrections.}

\end{deluxetable}



\clearpage

\clearpage
\clearpage

\clearpage

\end{document}